\documentclass{aa}  

\usepackage{graphicx}
\usepackage{txfonts}
\usepackage{lscape}
\usepackage{natbib,twoopt}
\bibpunct{(}{)}{;}{a}{}{,} 
\makeatletter
\newcommandtwoopt{\citeads}[3][][]{\href{http://adsabs.harvard.edu/abs/#3}%
        {\def\hyper@linkstart##1##2{}%
                \let\hyper@linkend\@empty\citealp[#1][#2]{#3}}}
\newcommandtwoopt{\citepads}[3][][]{\href{http://adsabs.harvard.edu/abs/#3}%
        {\def\hyper@linkstart##1##2{}%
                \let\hyper@linkend\@empty\citep[#1][#2]{#3}}}
\newcommandtwoopt{\citetads}[3][][]{\href{http://adsabs.harvard.edu/abs/#3}%
        {\def\hyper@linkstart##1##2{}%
                \let\hyper@linkend\@empty\citet[#1][#2]{#3}}}
\newcommandtwoopt{\citeyearads}[3][][]%
{\href{http://adsabs.harvard.edu/abs/#3}
        {\def\hyper@linkstart##1##2{}%
                \let\hyper@linkend\@empty\citeyear[#1][#2]{#3}}}
\makeatother

\def\cm#1{\ifmmode {\,{\rm cm^{-#1}}}                  
        \else \hbox{$\,${\rm cm$^{\rm -#1}$}}\fi}
\def\raw {\ifmmode\rightarrow\else$\rightarrow$\fi}
\def\ex#1{\ifmmode {\times 10^{#1}}         
        \else \hbox{{$\times 10^{\rm #1}$}}\fi}

\newcommand{\kms}{\mbox{km~s$^{-1}$}}
\newcommand{\bdens}{\mbox{g~cm$^{-3}$}}

\newcommand{\clam}{\mbox{$\lambda_{\mathrm{c}}$}}
\newcommand{\dlam}{\mbox{$\Delta\lambda$}}

\newcommand{\av}{\mbox{$A_{V}$}}

\newcommand{\hal}{\mbox{H$\alpha$}}

\newcommand{\nexp}{\mbox{$N_{\mathrm{exp}}$}} 
\newcommand{\npol}{\mbox{$N_{\mathrm{pol}}$}}
\newcommand{\ndith}{\mbox{$N_{\mathrm{dither}}$}} 

\newcommand{\mloss}{\mbox{$\dot{M}$}}
\newcommand{\my}{\mbox{$M_{\odot}$~yr$^{-1}$}}

\newcommand{\ls}{\mbox{$L_{\odot}$}}
\newcommand{\lbol}{\mbox{$L_{\mathrm{bol}}$}}

\newcommand{\fbol}{\mbox{$F_{\mathrm{bol}}$}}

\newcommand{\msun}{\mbox{$M_{\odot}$}}

\newcommand{\rstar}{\mbox{$R_{\star}$}}
\newcommand{\rs}{\mbox{$R_{\star}$}}
\newcommand{\rsp}{\mbox{$R_{\star}^{\prime}$}}

\newcommand{\teff}{\mbox{$T_{\mathrm{eff}}$}}

\newcommand{\tb}{\mbox{$T_{\mathrm{bright}}^{\prime}$}}

\newcommand{\vel}{\mbox{$\varv$}} 
 
\newcommand{\vinf}{\mbox{$\varv_{\infty}$}}

\newcommand{\vlsrsys}{\mbox{$V^{\mathrm{LSR}}_{\mathrm{sys}}$}} 

\newcommand{\microns}{\mbox{$\mu$m}}

\newcommand{\ik}{\object{IK~Tau}}
\newcommand{\wh}{\object{W~Hya}}
\newcommand{\hip}{\object{HIP~20188}}

\newcommand{\alum}{Al$_{2}$O$_{3}$}
\newcommand{\enst}{MgSiO$_{3}$}
\newcommand{\fors}{Mg$_{2}$SiO$_{4}$}

\newcommand{\tauV}{\mbox{$\tau_{550\,\mathrm{nm}}$}}

\begin{document}

        \title{Exploring the innermost dust formation region of the oxygen-rich AGB star \ik\ with VLT/SPHERE-ZIMPOL and VLTI/AMBER\thanks{Based on SPHERE and AMBER observations made with the Very Large Telescope and Very Large Telescope Interferometer of the European Southern Observatory. Programme ID: 098.D-0523(A) and 098.D-0523(C).}}
        
        \author{C. Adam\inst{1}\fnmsep\thanks{\email{christian.adam84@gmail.com}} \and K. Ohnaka\inst{1}}
        
        \institute{\inst{1} Universidad Cat\'{o}lica del Norte, Instituto de Astronom\'{\i}a, 0610 Avenida Angamos, Antofagasta, Chile
        }

        \date{Received / Accepted}
        
        
        \abstract
        {
                Low- and intermediate-mass stars at the asymptotic giant branch (AGB) are known to be prevalent dust providers to galaxies, replenishing the surrounding medium with molecules and dust grains. However, the mechanisms responsible for the formation and acceleration of dust in the cool extended atmospheres of AGB stars are still open to debate.
        }
        {
                We present visible polarimetric imaging observations of the oxygen-rich AGB star \ik\ obtained with the high-resolution polarimetric imager VLT/SPHERE-ZIMPOL at post-maximum light (phase 0.27) as well as high-spectral resolution long-baseline interferometric observations with the AMBER instrument at the Very Large Telescope Interferometer (VLTI). We aim to spatially resolve the dust and molecule formation regions, and to investigate their physical and chemical properties within a few stellar radii of \ik.
                
        }
        {
                \ik\ was observed with VLT/SPHERE-ZIMPOL at three wavelengths in the pseudo-continuum (645, 748, and 820\,nm), in the \hal\ line at 656.3\,nm, and in the TiO band at 717\,nm. The VLTI/AMBER observations were carried out in the wavelength region of the CO first overtone lines near 2.3\,\microns\ with a spectral resolution of 12\,000.

        }
        {
                The excellent polarimetric imaging capabilities of SPHERE-ZIMPOL have allowed us to spatially resolve clumpy dust clouds at 20--50\,mas from the central star, which corresponds to 2--5\,\rs\ when combined with a central star's angular diameter of 20.7$\pm$1.53\,mas measured with VLTI/AMBER. The diffuse, asymmetric dust emission extends out to $\sim$73\,\rs.
                We find that the TiO emission extends to 150\,mas (15\,\rs). The AMBER data in the individual CO lines also suggest a molecular outer atmosphere extending to $\sim$1.5\,\rs. The results of our 2-D Monte Carlo radiative transfer modelling of dust clumps suggest that the polarized intensity and degree of linear polarization can be reasonably explained by small-sized (0.1\,\microns) grains of \alum, \enst, or \fors\ in an optically thin shell (\tauV=0.5$\pm$0.1) with an inner and outer boundary radius of 3.5\,\rs\ and $\gtrsim$25\,\rs, respectively.
                The observed clumpy structures can be reproduced by a density enhancement of a factor of 3.0$\pm$0.5. However, the model still predicts the total intensity profiles to be too narrow compared to the observed data, which may be due to the TiO emission and/or grains other than homogeneous, filled spheres.
        }
        {       \ik's mass-loss rate is 20 to 50 times higher than the previously studied AGB stars \wh, \object{R~Dor}, and \object{$o$~Cet}. Nevertheless, our observations of \ik\ revealed that clumpy dust formation occurs close to the star as seen in those low mass-rate AGB stars.
        
        }
        
        \titlerunning{High resolution observations of \ik\ with ZIMPOL and AMBER}
        \authorrunning{C.~Adam \& K.~Ohnaka} 
        \maketitle
        %
        
        \begin{table*}
                \caption{\label{table:1}Summary of SPHERE-ZIMPOL observations for \ik\ and its PSF reference star \hip.} 
                \centering      
                \begin{tabular}{lccccccccccc}
                        \hline\hline
                        
                        \# & $t_{\mathrm{obs}}$ & DIT & NDIT & \nexp & \npol & \ndith & Filter & Seeing & AM & Strehl & Strehl \\    
                        & (UTC) & (s) & & & & & (cam1/cam2) & (\arcsec) & & ($H$) & (visible) \\[0.5mm]
                        \hline  
                        \multicolumn{12}{c}{\ik: 2016 Nov 20 (UTC)} \\
                        1 & 02:40:58 & 80.0 & 2 & 1 & 2 & 3 & CntHa/NHa & 0.62 & 1.38 & 0.74 & 0.14 \\ 
                        2 & 03:21:54 & 10.0 & 2 & 1 & 7 & 3 & TiO717/Cnt748 & 0.91 & 1.25 & 0.71 & 0.18 \\
                        3 & 06:57:36 & 5.0 & 4 & 1 & 3 & 3 & Cnt820/Cnt820 & 0.68 & 1.6 & 0.73 & 0.28\\
                        \hline 
                        \multicolumn{12}{c}{\hip: 2016 Nov 20 (UTC)} \\
                        C1 & 05:21:49 & 4.0 & 40 & 1 & 1 & 3 & CntHa/NHa & 0.75 & 1.23 & 0.77 & 0.22 \\  
                        C2 & 05:59:43 & 2.0 & 6 & 1 & 7 & 3 & TiO717/Cnt748 & 1.16 & 1.28 & 0.79 & 0.28 \\
                        C3 & 06:30:04 & 2.0 & 8 & 1 & 3 & 3 & Cnt820/Cnt820 & 0.76 & 1.34 & 0.82 & 0.41\\
                        \hline 
                \end{tabular}
                \tablefoot{DIT: detector integration time. NDIT: number of frames. \nexp: number of exposures of each polarization component in each polarization cycle at each dithering position. \npol: number of polarization cycles at each dithering position. \ndith: number of dithering positions. The seeing was measured in the visible. AM: airmass. The Strehl ratios in the visible were calculated from the $H$-band Strehl ratios for \ik, while they were measured from the observed ZIMPOL images for \hip.}
        \end{table*}
        
        \section{Introduction}
        
        It is well known from observations that low- and intermediate-mass stars (0.8\,\msun$<$\,M\,$<$\,8\,\msun) experience extensive mass loss during the late stages of their evolution, particularly throughout the asymptotic giant branch (AGB) phase. During this stage, material is stripped away from the stellar surface by low-velocity stellar winds (\vel$\sim$5--15\,\kms), at mass-loss rates between 10$^{-8}$ to 10$^{-4}$\,\my. It is considered that large-amplitude stellar pulsations lift material to heights where temperatures are sufficiently low to allow for dust grains to condense from the gas phase \citep{2003agbs.conf.....H,2018A&ARv..26....1H}. 
        
        It is thought that the wind acceleration is triggered by the radiation pressure on dust grains that they receive due to their absorption of stellar photons. And indeed, theoretical simulations based on this scenario are in good agreement with observations of carbon-rich AGB stars \citep[e.g.,][]{1991A&A...242L...1F,1992A&A...266..321F}. However, the situation is different in the case of oxygen-rich AGB stars where hydrodynamical simulations \citep{2006A&A...460L...9W,2007ASPC..378..145H} showed that the radiation pressure on the dust grains due to absorption is not sufficient to drive the mass loss at the rates observed.    
        Observations of oxygen-rich AGB stars \citep{2012Natur.484..220N,2016A&A...589A..91O, 2007A&A...470..191W,2013A&A...551A..72S,2013A&A...560A..75K}, on the other hand, point to the synthesis of dust grains, such as alumina and/or iron-free silicates, close to the star ($\sim$2--5\,\rs). Because alumina and iron-free silicate opacities are low in the wavelength range where most of the stellar flux is emitted, scattering on transparent, micron-sized (>\,0.1\,\microns) silicates grains was proposed as a viable mechanism to accelerate the wind in oxygen-rich AGBs at small radii \citep{2008A&A...491L...1H,2012A&A...546A..76B}. High angular resolution observations in the visible are therefore crucial to investigate the distribution of these grains in the immediate surrounding of AGB stars and to constrain their influence on the dust formation and wind acceleration processes.
        
        The Mira-type star \ik, also known as \object{NML~Tau}, is one of the best studied oxygen-rich AGB-stars and often considered a reference of its class. \ik\ was discovered in 1965 by \cite{1965ApJ...142..399N} and is an extremely red Mira-type variable with brightness variations of $\Delta V\sim$\,6\,mag in the $V$-band over a period of $\sim$460 days \citep{2018A&A...612A..48W}. Consequently, its spectral type varies from M8 to M11 \citep{2009yCat.1280....0K}. \cite{1997ApJ...490..407H} deduced a distance of 265\,pc from dust shell motions detected at 11\,\microns\ with the Infrared Spatial Interferometer (ISI), similar to the 266\,pc estimated by \cite{2012A&A...546A..16R}, based on the investigation of 22-GHz water maser clouds. 
        The recently published second data release of the GAIA mission (Gaia DR2) locates \ik\ slightly farther away at a distance of $D$\,=\,284$^{+29}_{-36}$\,pc \citep{2018A&A...616A...1G}. We adopted an average distance of $D$=280$\pm$30\,pc for our calculations taking the different distance measurements into account. 
        \ik\ is also a high mass-loss rate AGB star with estimated mass-loss rates ranging from $\sim$3.8$\times$10$^{-6}$\,\my\ \citep{1998A&AS..130....1N} up to $\sim$4.5$\times$10$^{-6}$\,\my, or \mloss\,$\sim$4.7$\times$10$^{-6}$\,\my\ as measured by \cite{2010A&A...516A..69D} and \cite{2010A&A...516A..68K}, respectively. The systemic velocity of the star with respect to the local standard of rest is \vlsrsys\, $\sim$34\,\kms\ \citep[][and references therein]{2010A&A...516A..68K} and the terminal wind expansion velocity \vinf\ ranges from 17.7 to 18.5\,\kms\ \citep{2018A&A...615A..28D,2016A&A...591A..44M,2013A&A...558A.132D}. 
        The oxygen-rich circumstellar envelope (CSE) around \ik\ also displays maser emission of OH, H$_{2}$O, and SiO \citep{1987ApJ...323..756L,1989ApJ...340..479B,2010A&A...516A..68K} and so far over 34 different molecular species \citep[including different isotopologues,][]{2018A&A...615A..28D} have been identified in \ik\ including molecules such as SiO, AlO, AlOH, TiO, or TiO$_{2,}$ which are considered to be the gaseous precursors of dust grains like alumina (\alum), or silicates such as enstatite (\enst), or forsterite (\fors). 
        Although \alum\ dust is stable at very high temperatures, gaseous \alum\ has a low abundance, calling into question its role as the first candidate to form condensates. For a long time, TiO and TiO$_{2}$ were considered to be the best candidates as primary condensates. However, more recent ALMA observations of TiO and TiO$_{2}$ transitions in \ik, \object{R~Dor}, or Mira \citep{2018A&A...615A..28D,2017A&A...599A..59K} show these two molecules are abundant in the gas phase well beyond the dust formation regions, indicating that titanium might not to be efficiently depleted from the gas phase around O-rich stars, which would imply that titanium oxides may not be important first condensates as previously thought \citep{2017A&A...599A..59K}.
        
        In this paper, we aim to compare the first visible light polarimetric observations in high angular resolution of \ik\ with 2-D radiative transfer modelling to probe the dust-formation and wind-driving processes and spatially resolve the wind-acceleration regions in the immediate vicinity of \ik. 
        In Sect.~\ref{sec:obs}, we describe the observations obtained with the Spectro-Polarimetric High-contrast Exoplanet REsearch \citep[SPHERE;][]{2008SPIE.7014E..18B} instrument and with the Astronomical Multi-BEam combineR \citep[AMBER;][]{2007A&A...464....1P} instrument at the Very Large Telescope Interferometer (VLTI), followed by the observational results presented in Sect.~\ref{sec:obs_res}. The determination of the effective temperature and luminosity is described in Sect.~\ref{sec:stell_par}. Furthermore, we describe the radiative transfer modelling of the data, its results and interpretation in Sect.~\ref{sec:rtm}, followed by final conclusions presented in Sect.~\ref{sec:concl}.
        
        \section{Observations and data reduction}\label{sec:obs}
        
        \subsection{SPHERE/ZIMPOL visible polarimetric imaging} \label{subsec:obs_zimpol}
        
        The Zurich IMaging Polarimeter \citep[ZIMPOL;][]{2008SPIE.7014E..3FT} is one of the focal plane subsystems within the SPHERE instrument. In combination with the extreme adaptive optics (AO) system SAXO \citep{2006OExpr..14.7515F}, SPHERE-ZIMPOL provides high contrast and high spatial resolution polarimetric observations in the wavelength range from 550\,nm to 900\,nm.
        
        Our SPHERE-ZIMPOL observations of \ik\ took place on 2016 November 20 (UTC, Programme ID: 098.D-0523(A/B), P.I.: K. Ohnaka). We used the light curve of \ik\ of the American Association of Variable Star Observers (AAVSO) in order to have an estimate of the $V$ magnitude at the time of our observations. We find a $V$ magnitude of 12.6\,$\pm$\,0.2\,mag that corresponds to phase 0.27 (post-maximum light).
        In addition to the science target \ik, we observed the K0 star \hip\ as a point spread function (PSF) reference star using the same instrumental set-up as for \ik. According to the CalVin database\footnote{http://www.eso.org/observing/etc/bin/gen/form?INS.NAME=\newline CALVIN+INS.MODE=CFP} \citep{2016A&A...589A.112C}, \hip\ has an angular diameter of 1.265$\pm$0.111\,\mbox{mas} and, thus should appear as a point source at the spatial resolution (20--30\,\mbox{mas}) of the SPHERE-ZIMPOL instrument.
        
        We used ZIMPOL in field tracking, polarimetric (P2) mode, employing five different filters: CntHa with a central wavelength \clam=644.9\,nm and a full-width at half maximum (FWHM) \dlam=4.1\,nm; NHa with \clam=656.34\,nm and \dlam=0.97\,nm; TiO717 with \clam=716.8\,nm and \dlam=19.7\,nm; Cnt748 with \clam=747.4\,nm and \dlam=20.6\,nm; as well as the Cnt820 filter with \clam=817.3\,nm and \dlam=19.8\,nm. ZIMPOL is equipped with two camera arms, camera 1 and camera 2, each with a pixel scale of 3.628\,mas, providing a field of view (FoV) of about 3.5\,\arcsec$\times$\,3.5\,\arcsec. The data are always taken simultaneously in both arms, each equipped with its own filter wheel, which allows the simultaneous selection of two identical or different filters. We used the filter pairs (CntHa, NHa), and (TiO717, Cnt748) to probe \hal\ emission and TiO emission, as well as (Cnt820, Cnt820). For both targets and with each filter pair, we took \nexp\ exposures for each of the Stokes $Q^{+}$, $Q^{-}$, $U^{+}$, and $U^{-}$ components, with  NDIT (number of detector integrations) frames in each exposure. The polarization cycle of $Q^{+}$, $Q^{-}$, $U^{+}$, and $U^{-}$ was then repeated \npol\ times. This procedure (\nexp\,$\times$\,\npol\ exposures) was carried out at three different dithering positions. The summary of our SPHERE-ZIMPOL observations is presented in Table~\ref{table:1}. 
        
        We also list observing conditions at the time of observation, such as the average seeing, and airmass (AM) as well as the Strehl ratios in the $H$-band and the visible.
        The $H$-band Strehl ratios are recorded by SPHERE in separate FITS files\footnote{``Classified as OBJECT, AO'' in the ESO data archive}, the so-called GEN-SPARTA data.
        For \ik\ the median $H$-band Strehl ratios during the observations was 0.71--0.74, while the observed median $H$-band Strehl ratio for \hip\ is 0.77--0.82.
        
        We measured the Strehl ratio in the visible for the PSF reference \hip\ directly from the total intensity maps. Due to its angular extension we cannot measure the Strehl ratio directly for \ik\ as opposed to \hip,\ which appears as point source. However, we still can estimate the Strehl ratio in the visible from the $H$-band-Strehl ratio exploiting the fact that the Strehl ratio can be approximated using the Mar\'{e}chal approximation \citep[see e.g.][Sect. 5.4.4]{2000plbs.conf.....L}:
        \begin{eqnarray}
                \mathrm{S} &=& \exp(-\sigma_{\phi}^{2}) \, , 
        \end{eqnarray}
        where $\sigma_{\phi}^2$ is the phase variance in radian$^2$. Since  $\sigma_{\phi}$ is proportional to $1/\lambda$, the Strehl ratio  $\mathrm{S}_{\lambda}$ at any given wavelength can then be estimated via
        \begin{eqnarray}
        \mathrm{S}_{\lambda} &=& \exp \left[\left( \frac{\lambda_{H}}{\lambda}\right)^2 \cdot \ln \mathrm{S}_{H} \right]  \, . 
        \end{eqnarray}
        
        We tested this method on \hip. The calculated Strehl ratios in the visible 0.18 (CntHa/NHa), 0.30 (TiO717/Cnt748), and 0.45 (Cnt820/Cnt820) are in good agreement with the directly measured Strehl ratios of the PSF reference \hip, 0.22 (CntHa/NHa), 0.28 (TiO717/Cnt748), and 0.41 (Cnt820/Cnt820). Applying this method to \ik\ we derive Strehl ratios of about 0.14 (CntHa/NHa), 0.18 (TiO717/Cnt748), and 0.28 (Cnt820/Cnt820).
        
        For the reduction of our data we used the SPHERE pipeline version 0.36.0\footnote{Available at ftp://ftp.eso.org/pub/dfs/pipelines/sphere}. We processed each exposure with the pipeline, which as a result produces the de-rotated image of the $Q^{+}$, $Q^{-}$, $U^{+}$, and $U^{-}$ component as well as the corresponding intensity component, namely $I_{Q^{+}}$, $I_{Q^{-}}$, $I_{U^{+}}$, and $I_{U^{-}}$, respectively, for each camera. The output images were combined and averaged to produce the final images of the polarization and there associated intensity. From this we calculated the final Stokes parameter $Q$ and $U$ via 
        \begin{eqnarray}
        I_{Q} &=& 0.5 \cdot \left(I_{Q^{+}} + I_{Q^{-}}\right) ,\\
        Q     &=& 0.5 \cdot \left(Q^{+} - Q^{-}\right),\;\\
        I_{U} &=& 0.5 \cdot \left(I_{U^{+}} + I_{U^{-}}\right) ,\\
        U     &=& 0.5 \cdot \left(U^{+} - U^{-}\right),\;
        \end{eqnarray}
        as well as the total intensity $I$ employing
        \begin{eqnarray}
        I &=& 0.5 \cdot \left(I_{Q} + I_{U}\right) \, . 
        \end{eqnarray}
        We then computed the polarized intensity $I_{\mathrm{p}}$, the degree of linear polarization $p_{\mathrm{L}}$ and the position angle $\theta$ of the polarization vector from the Stokes parameter via
        \begin{displaymath}
        I_{\mathrm{p}} = \sqrt{Q^{2} + U^{2}}\, , \;
        p_{\mathrm{L}} = I_{\mathrm{p}} / I \, , \;
        \theta = 0.5\cdot\arctan(U/Q)\, .
        \end{displaymath}
        
        The only spectrum of \ik\ in the wavelength range of our SPHERE observations available in the literature is the high-resolution echelle, long-slit spectrum taken by \citet{2008ApJS..179..166S}. However, given the time difference between their observations and ours we did not use this spectra to flux-calibrate our data of \ik\ at the time of our observations.
        
        We therefore flux-calibrated the intensity maps of \ik\ following the approach presented by \cite{2017A&A...597A..20O}, employing the PSF reference star to calibrate the flux of our SPHERE-ZIMPOL observations of \ik. We used the Library of Stellar Spectra \citep{1998PASP..110..863P} to approximate the spectrum of \hip\ with a template spectrum whose effective temperature and spectral classification are closest to that of the reference star. The K-type star \hip\ has an effective temperature of \teff=4173\,K \citep[Gaia DR2,][]{2018A&A...616A...1G}. Based on these data we adopted the spectrum of a K3--4II star available in the Library of Stellar Spectra. After that, and in order to account for the interstellar extinction \av\ at the wavelengths of our observations with ZIMPOL, we reddened the template spectrum, using the parametrization of the interstellar extinction by \cite{1992A&A...258..104A} and its wavelength-dependence published by \cite{1989ApJ...345..245C}. For the PSF reference star \hip\ we find \av=0.338$\pm$0.206\,mag, adopting a distance of $D$$=$196.7$\pm$3.5\,pc \citep[Gaia DR2,][]{2016A&A...595A...1G,2018A&A...616A...1G}. The reddened spectrum could then be scaled so that the computed flux, within a circular aperture of radius 1.5\arcsec, is equal to a $V$-band magnitude of 6.3\,mag of \hip\ \citep{2009yCat.1280....0K}. The aperture was chosen so that it contains as much flux from the star as possible, but also stays well within the detector. We then used the flux-calibrated template spectrum to compute the flux of PSF reference star with each ZIMPOL filter, in which we approximated the transmission curves of each filter with a top-hat function specified by the central wavelength and FWHM. With these fluxes we were finally able to flux-calibrate the images of \ik, taking into account the uncertainties in distance and interstellar extinction for the reference star and \ik. The derived fluxes and relative uncertainties with the five SPHERE-ZIMPOL filter are listed in Table~\ref{table:2}. We find relative errors of about 3\% to 8\% in the different filter.
        
        \begin{table}[hbtp!] 
                \caption{Flux of \ik\ derived for five filters of our SPHERE-ZIMPOL observations.} 
                \label{table:2} 
                \centering      
                \begin{tabular}{l c c c c}
                        \hline\hline\\[-1em]
                        
                        Filter & \clam & FWHM & Flux & relative \\
                         & & & (phase 0.27) & flux error \\
                        & (nm) & (nm) & $\mathrm{(W\,m^{-2}\,\mu m^{-1})}$ & (\%) \\
                        \hline
                        CntHa & 644.9 & 4.1 & 4.58$\times$10$^{-12}$ & 2.9\\ 
                        NHa & 656.34 &0.97& 7.03$\times$10$^{-12}$ & 3.3 \\
                        TiO717 & 716.8 &19.7 & 1.07$\times$10$^{-11}$ & 5.0 \\
                        Cnt748 & 747.4 & 20.6 & 2.91$\times$10$^{-11}$ & 5.9 \\
                        Cnt820 & 817.3 & 19.8 & 1.38$\times$10$^{-10}$ & 7.9 \\
                        \hline 
                \end{tabular}
        \end{table}
        
        \begin{table*}[hbtp!] 
                \caption {
                        Summary of VLTI/AMBER observations of \ik\ and calibrator \object{Rigel}. 
                }
                \begin{center}
                        
                        \begin{tabular}{l c c c c c l }\hline\hline
                                \# & $t_{\rm obs}$ & $B_{\rm p}$ & PA & Seeing & $\tau_0$ &
                                ${\rm DIT}\times{\rm N}_{\rm f}\times{\rm N}_{\rm exp}$ \\ 
                                & (UTC) & (m) & (\degr) & (\arcsec) & (ms) & (sec) \\
                                \hline
                                \multicolumn{7}{c}{\ik: 2016 December 19 (UTC)}\\
                                \hline
                                1 & 05:02:35 & 11.1/15.2/16.6 & 36/$-67$/$-27$ & 0.48 & 8.6 & 1$\times$70$\times$5 \\
                                \hline
                                \multicolumn{7}{c}{\object{Rigel}: 2016 December 19 (UTC)}\\
                                \hline
                                C1 & 04:36:34 & 11.1/21.8/24.0 & 31/$-63$/$-35$ & 0.93 & 6.0 &1$\times$70$\times$5\\
                                C2 & 05:22:52 & 11.3/20.4/23.0 & 33/$-59$/$-30$ & 0.47 & 9.5 &1$\times$70$\times$5\\
                                \hline
                                \label{obs_log_amber}
                                \vspace*{-7mm}
                                
                        \end{tabular}
                \end{center}
                \tablefoot{
                        $B_{\rm p}$: Projected baseline length. PA: Position angle of the baseline 
                        vector projected onto the sky. DIT: Detector Integration Time. $N_{\rm f}$: Number of frames in each exposure. $N_{\rm exp}$: Number of exposures. The seeing and the coherence time ($\tau_0$) were measured in the visible. 
                }
        \end{table*}
        
        As mentioned above, \ik\ is relatively faint in the V-band, with about 12.6\,mag, compared to the reference star \hip, with a $V$-band magnitude of 6.3\,mag. To test how far this difference influences the performance of the adaptive optics, we also estimate the brightness of \ik\ and the PSF reference star \hip\ as seen by the wavefront sensor (WFS) in the AO system of SPHERE. The WFS of SPHERE has a central wavelength of 800--850\,nm (J. Milli, priv. comm.). We therefore use the fluxes derived for the Cnt820 filter (\clam=817.3$\pm$19.8\,nm) to get an estimate on the I-band magnitude of \ik. For a flux of about 1.4$\times$10$^{-10}\,\mathrm{W\,m^{-2}\,\mu m^{-1}}$ from the Cnt820 filter, and applying the zero-point flux density derived by \citet{1998A&A...333..231B}, we estimate an I-band magnitude of $I$\,$\approxeq$4.8\,mag for \ik. For the PSF reference we find an I-band magnitude of 4.58\,mag \citep{2014ASPC..485..223B}. The small magnitude difference therefore should not have affected the performance of the AO significantly.

        \subsection{Long-baseline spectro-interferometric observation with VLTI/AMBER}
        \label{subsect_obs_amber}
        
        We observed \ik\ with the near-infrared interferometric instrument AMBER \citep{2007A&A...464....1P} at VLTI to measure its angular diameter, which is needed for the radiative transfer modelling presented below. Our VLTI/AMBER observation of \ik\ was carried out on 2016 December 19 (UTC), a month after the SPHERE-ZIMPOL observations (Programme ID: 098.D-0523(C), P.I.: K.~Ohnaka). The Auxiliary Telescope (AT) configuration A0-B2-C1 resulted in projected baselines of 11.1, 15.2, and 16.6\,m. We used the high spectral resolution of 12\,000, covering from 2.26 to 2.31\,\mbox{$\mu$m}, with the fringe tracker FINITO. We assumed that the time variation in the angular diameter between the SPHERE-ZIMPOL observations and the AMBER observations does not affect our modelling, given that the difference in the variability phase is just 0.078.
        We observed \object{Rigel} 
        \citep[\object{$\beta$~Ori}, B8Iae, $K$=0.18, uniform-disk 
        diameter = 2.69$\pm$0.36\,mas,][]{2017yCat.2346....0B} as an interferometric and spectroscopic calibrator. 
        A summary of our AMBER observations is given in Table~\ref{obs_log_amber}. 
        
        The AMBER data were reduced using the amdlib ver~3.0.8\footnote{Available at http://www.jmmc.fr/data\_processing\_amber.htm}, which is based on the P2VM algorithm \citep{2007A&A...464...29T,2009A&A...502..705C}. 
        The amdlib software derives the (uncalibrated) squared visibility amplitude, closure phase (CP), and differential phase (DP) from the recorded interferograms for the science target and the calibrator. Then we derived the calibrated interferometric observables by taking the best 20\% and 80\% of the frames in terms of the fringe signal-to-noise ratio. 
        In the present paper, the visibility amplitude obtained with the best 20\% of the frames is used, because the errors are smaller than with the best 80\%. 
        On the other hand, we present the CPs and DPs obtained with the best 80\% 
        because of their smaller errors. Details of the reduction are described in Ohnaka et al. (\citeyear{2009A&A...503..183O}, \citeyear{2011A&A...529A.163O}, and \citeyear{2013A&A...555A..24O}). 
        The wavelength calibration was carried out using the telluric lines with the method described in \cite{2013A&A...555A..24O}. 
        The spectroscopic calibration of \ik\ was done by dividing 
        the observed spectrum of \ik\ with that of \object{Rigel}, because the true spectrum 
        of \object{Rigel} is not expected to show any spectral features in the observed wavelength range at the spectral resolution of 12\,000. We confirmed this by using the high-resolution ($\lambda/\Delta \lambda \approx 45\,000$) spectrum of \object{HD~87737} (A0Ib) available in the IGRINS spectral library\footnote{http://starformation.khu.ac.kr/IGRINS\_spectral\_library} \citep{2018ApJS..238...29P}. The star \object{HD~87737} has the closest spectral type and luminosity class to \object{Rigel} in the IGRINS infrared stellar spectral library. Its spectrum convolved down to AMBER's spectral resolution of 12\,000 does not show stellar spectral features, which justifies the use of \object{Rigel} as a featureless spectroscopic standard star for the observed spectral window.
        
        \begin{figure*}[hbtp!] 
                \includegraphics[width=175mm]{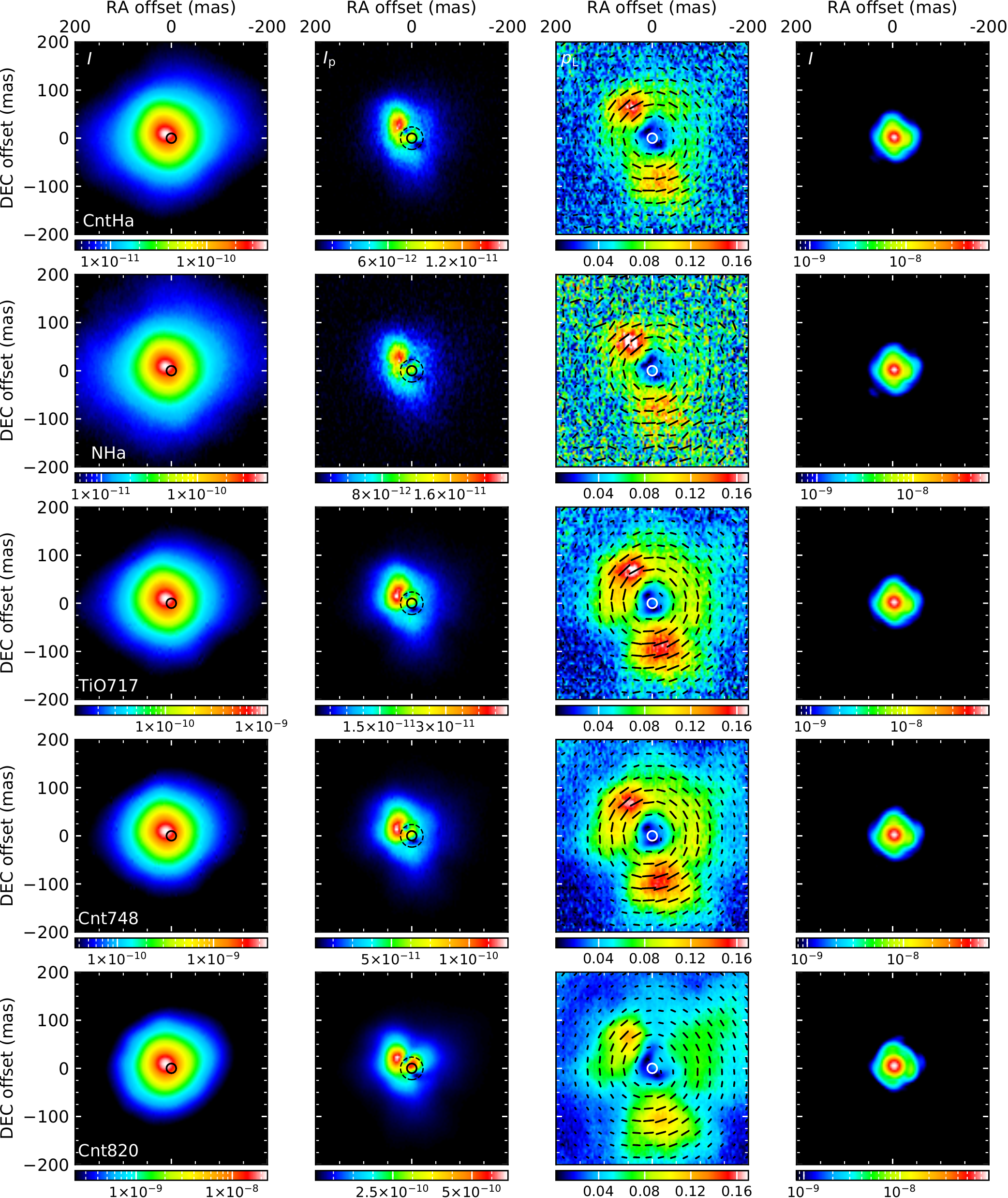}
                \caption{SPHERE-ZIMPOL polarimetric imaging observations of \ik. {\bf Left to right:} Each row shows the total intensity (first column), polarized intensity (second column), in units of \mbox{$\mathrm{W\,m^{-2}\,\mu m^{-1}\,arcsec^{-2}}$} as described in Sec.~\ref{subsec:obs_zimpol}, the degree of polarization superimposed with the polarization vector maps (third column), and the intensity of the PSF reference star \hip\ (last column), which are also presented in units of \mbox{$\mathrm{W\,m^{-2}\,\mu m^{-1}\,arcsec^{-2}}$}. {\bf Top to bottom:} Shown are the observed images at 645\,nm (CntHa, continuum), 656.3\,nm (NHa, \hal), 717\,nm (TiO717, TiO band), 748\,nm (Cnt748, continuum), as well as 820\,nm (Cnt820, continuum, average combined image of cam1 and cam2). The position and angular diameter of the star are indicated by the circles in columns one to three, whereas the approximate extension of the low-polarization region (radius $\sim$23\,mas, column three) is shown in column two as a dashed circle. The colour scale of the intensity maps of \ik\ and \hip\ is cut off at 0.5\% of the peak intensity. North is to the top and east to left in all panels.}
                \label{fig:1}
        \end{figure*}
        
        \begin{figure}[hbtp!] 
                \includegraphics[width=88mm]{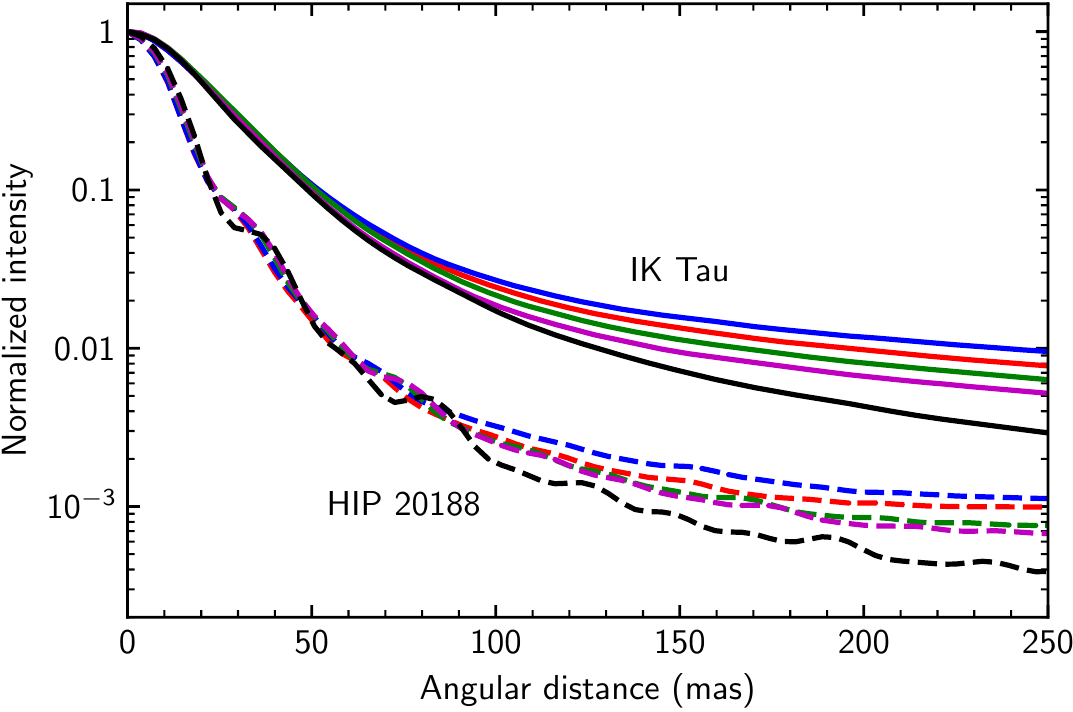}
                \caption{Azimuthally averaged 1-D intensity profiles of \ik\ (top, solid lines) and the PSF reference \hip\ (bottom, dashed lines) observed at at 645\,nm (red), 656.3\,nm (blue), 717\,nm (green), 748\,nm (magenta), and at 820\,nm (black), respectively.}
                \label{fig:2}
        \end{figure}
        
        \begin{figure*}[htbp!] 
                \includegraphics[width=180mm]{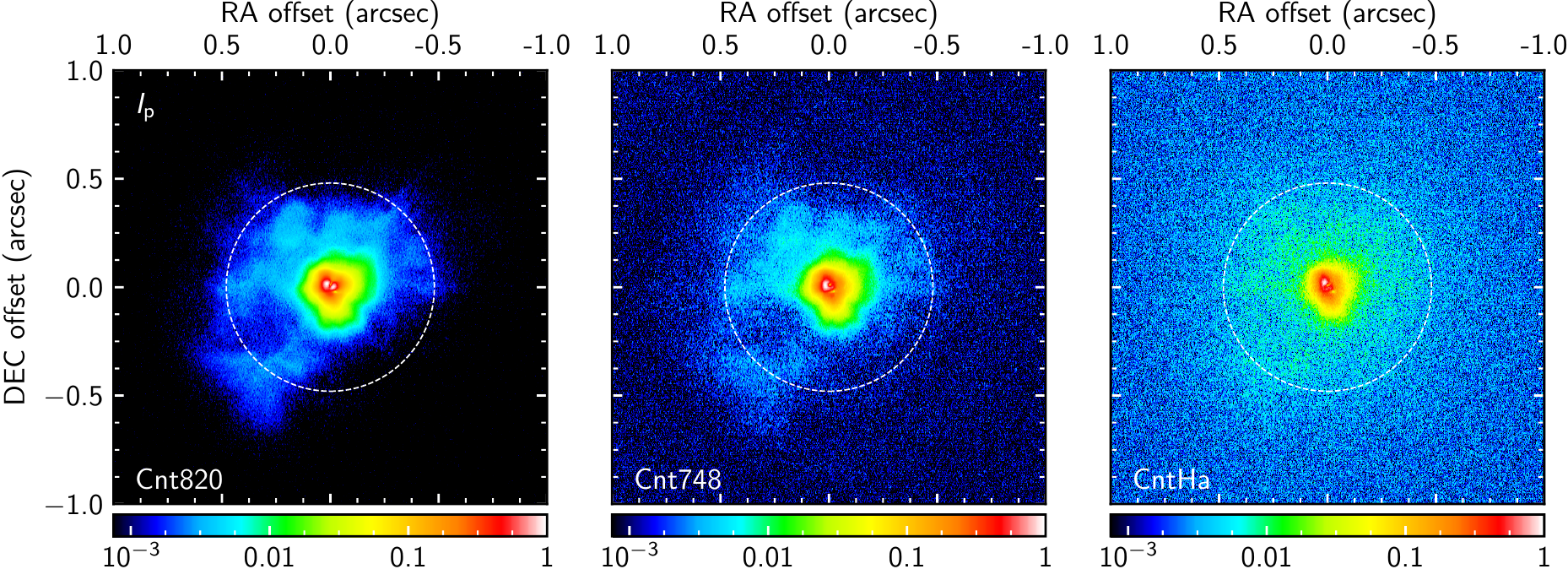}
                \caption{Extended structure in polarized intensity maps of \ik\ observed at 820\,nm (Cnt820), 748\,nm (Cnt748), and 645\,nm (CntHa). The approximate size AO correction ring (radius\,$\sim$480\,mas) is indicated by the (dashed) circle in each panel.}
                \label{fig:pl_extend}
        \end{figure*}
        
        \section{Observational results} \label{sec:obs_res}

        \subsection{Clumpy dust clouds and extended gas emission}
        \label{subsec:dust}
        
        Figure~\ref{fig:1} shows the total intensity $I$, polarized intensity $I_{\mathrm{p}}$, and the degree of polarization $p_{\mathrm{L}}$ maps (columns one to three) observed with the five ZIMPOL filters. Also shown, in the last column, are the intensity maps of the PSF reference star \hip. 
                
        The intensity profile of \ik\ appears noticeably more extended than that of the PSF reference, which is also demonstrated in the azimuthally averaged 1-D intensity profiles presented in Fig.~\ref{fig:2}. Applying a 2-D Gaussian fit to the images of \hip\ yields PSF FWHM of 26$\,\times$\,24\,mas at 645 and 656.3\,nm, 27\,$\times$\,26\,mas at 717 and 748\,nm, and 26\,$\times$\,25\,mas at 820\,nm. We find the FWHM of the intensity map \ik\ with 55$\,\times$\,55\,mas at 645\,nm, 56$\,\times$\,56\,mas at 656.3\,nm, 57\,$\times$\,53\,mas at 717\,nm, 54\,$\times$\,51\,mas at 748\,nm, and 53\,$\times$\,52\,mas at 820\,nm indicating that the CSE has been spatially resolved. 
        As can be seen in Table~\ref{table:1}, the Strehl ratio for \ik\ are slightly lower than for the PSF reference star. However, since they are comparable, therefore, it is unlikely that the extended emission of \ik\ is only due to the lower AO performance.
        
        In general the total intensity appears to be spherical in all observed filters.
        The peak of the intensity, however, is slightly displaced towards the north-east (NE) with respect to the centre of the central low polarization region in the degree of polarization maps (third column of Fig.~\ref{fig:1}), which we assume to be the location of the central star. The reason for this displacement could be a bright dust clump or inhomogeneities in the atmosphere such as a bright spot, which can be interpreted as fluctuations in the density and/or molecular abundance and/or excitation \citep{2016A&A...591A..70K}. The total intensity map, however, only shows the global extended structure of the circumstellar envelope.
        
        The polarized intensity maps, shown in the second column of Fig.~\ref{fig:1}, reveal more detailed insight in the innermost regions of the CSE. The central region of the polarized intensity is slightly elongated with a radius of about 100\,mas to 120\,mas. Given an angular diameter of $\theta_{\mathrm{D}}$=20.7$\pm$1.5\,mas derived from our VLTI/AMBER observations (see Sect.~\ref{sec:obs_res}), this corresponds to about 10--12\,\rstar. 
        
        We have estimated the errors per pixel in the polarized intensity using the output of the SPHERE pipeline, and here in particular the so-called RMS error maps. For the bright spot NE of the centre we find relative errors of 27\%, 34\%, 21\%, 18\%, and 13\% at 645\,nm, 656.3\,nm, 717\,nm, 748\,nm, and 820\,nm, respectively. The relative errors in the clump detected south of the central star are similar with 22\%, 40\%, 20\%, 19\%, and 16\% at 645\,nm, 656.3\,nm, 717\,nm, 748\,nm, and 820\,nm, respectively, while we find 27\%, 34\%, 21\%, 18\%, and 13\% in the central region.
        
        We detected two clumps that are visible in all five filters. One clump is located NE of the centre, with a peak at around 25\,mas (2.5\,\rstar) away from the centre. The second clump is located south of the central star at 30 to 50\,mas (3--5\,\rstar) and extending up to at least 120\,mas (12\,\rstar). 
        A third clump covering the central regions and apparently extending to the west up to about 75\,mas (7.5\,\rstar) was detected at 820\,nm. Marginal evidence of this clump-like structure can also be found in the other wavelengths observed, but without further observations we cannot rule out the possibility that this may be an artefact resulting from the reduction process.        
        
        The extension of dust formation region and the clumpy dust clouds in the CSE we observed are as well detected in other nearby AGB stars, such as images of the carbon-rich AGB star IRC+10216 \citep{1998A&A...333L..51W,1998A&A...334L...5H,2016MNRAS.455.3102S} or CIT6 \citep{2000ApJ...545..957M}. 
        Moreover, the polarimetric observations of \ik\ have revealed for the first time clumpy dust clouds forming close to the star in a high mass-loss, oxygen-rich AGB star with mass-loss rates 20 to 50 times higher than that of the low mass-loss rate, oxygen-rich AGB stars \object{R~Dor}, \wh\ and \object{$o$~Cet} \citep{2016A&A...591A..70K,2016A&A...589A..91O,2017A&A...597A..20O,2018A&A...620A..75K}, recently observed with SPHERE-ZIMPOL and studied in a similar way. 
        3-D hydrodynamical simulations suggest the formation of clumps to be a direct consequence of convection and/or stellar pulsations \citep{2008A&A...483..571F}. This provides a natural explanation for the observation of these structures made in \ik,\ as well as in the CSE of other O-rich AGB stars.

        The polarized intensity maps also reveal an extended structure in the outer regions of \ik\ as shown in Fig.~\ref{fig:pl_extend}. The inner region extends up to about 250\,mas (25\,\rstar), while the asymmetric diffuse emission reaches as far as about 700\,mas (70\,\rstar) to the south-east and about 400\,mas (40\,\rstar) to the north. 
        
        As mentioned earlier \ik\ shows strong SiO, H$_{2}$O, and OH masers emission. For example, \citet{2005ApJ...625..978B}  observed 43\,GHz SiO maser emission toward \ik\ using the Very Long Baseline Array (VLBA). Their results show a rotation axis in the north-east-south-west (NE-SW) direction. Our SPHERE-ZIMPOL data show no signature of an axis along NE-SW, with one dust clump towards NE, but the second clump clearly extends southwards. It therefore appears more likely that dust formation is governed by convection and/or pulsation, not directly by rotation. 
        
        The third column of Fig.~\ref{fig:1} shows the degree of linear polarization overlaid with the vector map of the position angle of polarization. The shell-like (though elongated) structure seen in the polarized intensity is also seen in the degree of linear polarization, and additionally supported by the concentric vector maps of the position angle of the degree of polarization. We measure a maximum degree of linear polarization of 16\%, 18\%, 16\%, 16\%, and 12\% at 645\,nm, 656.3\,nm, 717\,nm, 748\,nm, and 820\,nm, respectively. For the absolute errors on average measured over the clumps we find 0.05\% at 645\,nm, that is 16$\pm$0.05\%, 0.15\%, 0.1\%, 0.05\%, and 0.02\% at 656.3\,nm, 717\,nm, 748\,nm, and 820\,nm, respectively. Whereas in the centre the absolute errors are smaller with 0.004\%, 0.02\%, 0.02\%, 0.006\%, and 0.001\% at 656.3\,nm, 717\,nm, 748\,nm, and 820\,nm, respectively.
        
        The central region of very low polarization measures about 20--25\,mas in radius, which corresponds to 2 to 2.5\,\rs. 
        Within this region, the polarization signal is so low that the features in the polarized intensity and degree of polarization maps are not reliable. Therefore, we cannot confirm the presence or absence of dust in the region within 2--2.5\,\rs. 

        \subsection{\hal\ and TiO emission}\label{subsec:emmission}
        
        As mentioned in Section~\ref{subsec:obs_zimpol}, it is possible to observe two different filters, such as (CntHa, NHa) and (TiO717, Cnt748), simultaneously with SPHERE-ZIMPOL having the same AO performance for both filters. This allows us to investigate the presence of \hal\ emission and a more detailed study of TiO emission. 
        
        We first subtracted the CntHa image from the \hal\ image using the intensity maps of \ik\ (see Fig.~\ref{fig:1}, left column), which we have flux-calibrated as described in Sect.~\ref{subsec:obs_zimpol}. The results however are not conclusive for the following reasons. The 1-D azimuthal average profiles of the PSF reference star in general, show the same behaviour as the profile of \ik, as can be seen in Fig.~\ref{fig:2}. It is therefore possible that the observed flux difference is due to an unidentified instrumental effect that causes the source to appear more extended in the NHa filter than in the CntHa filter. It is also possible that a substantial fraction of the flux in the NHa filter might be completely unrelated to the \hal\ line, caused by more dominant molecular bands such as the TiO band, as observed in the visible spectrum presented by \cite{2008ApJS..179..166S} for \ik. Only visible-light spectra obtained simultaneously to the observations will allow the proper determination of whether or not the observed flux difference between the NHa and CntHa images corresponds to \hal\ emission.

        \begin{figure}[tbhp!] 
                \centering
                \includegraphics[width=70mm]{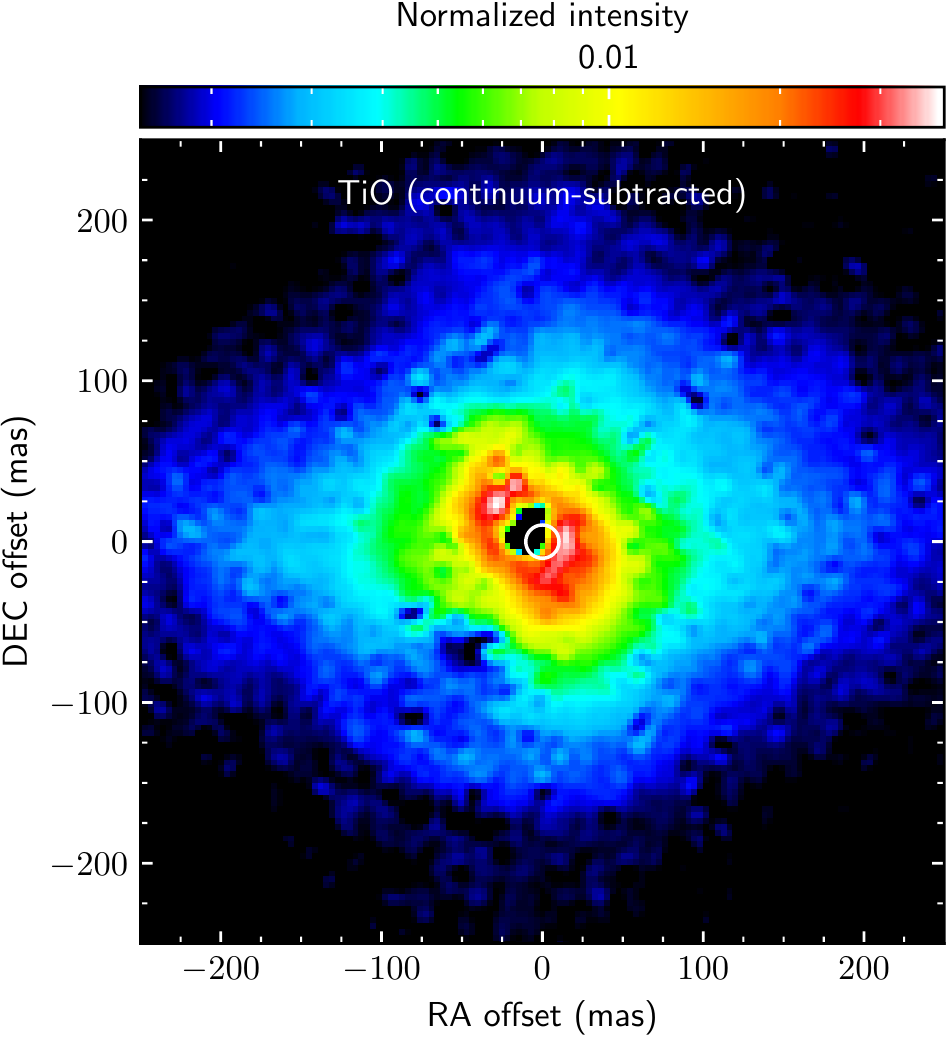}
                \caption{Continuum-subtracted TiO717 (716.8\,nm) image of \ik\ obtained form our SPHERE-ZIMPOL observations at post-maximum light (phase 0.27). The assumed angular diameter of the star ($\theta_{\mathrm{D}}$$\approx$21\,mas) as well as its approximate position are indicated by the white circle. Regions with negative pixel values are shown in black. North is to the top and east to left.}
                \label{fig:cont_subtr}
        \end{figure}
        
        \begin{figure}[hbtp!] 
                \begin{center}
                        \resizebox{\hsize}{!}{\rotatebox{0}{\includegraphics{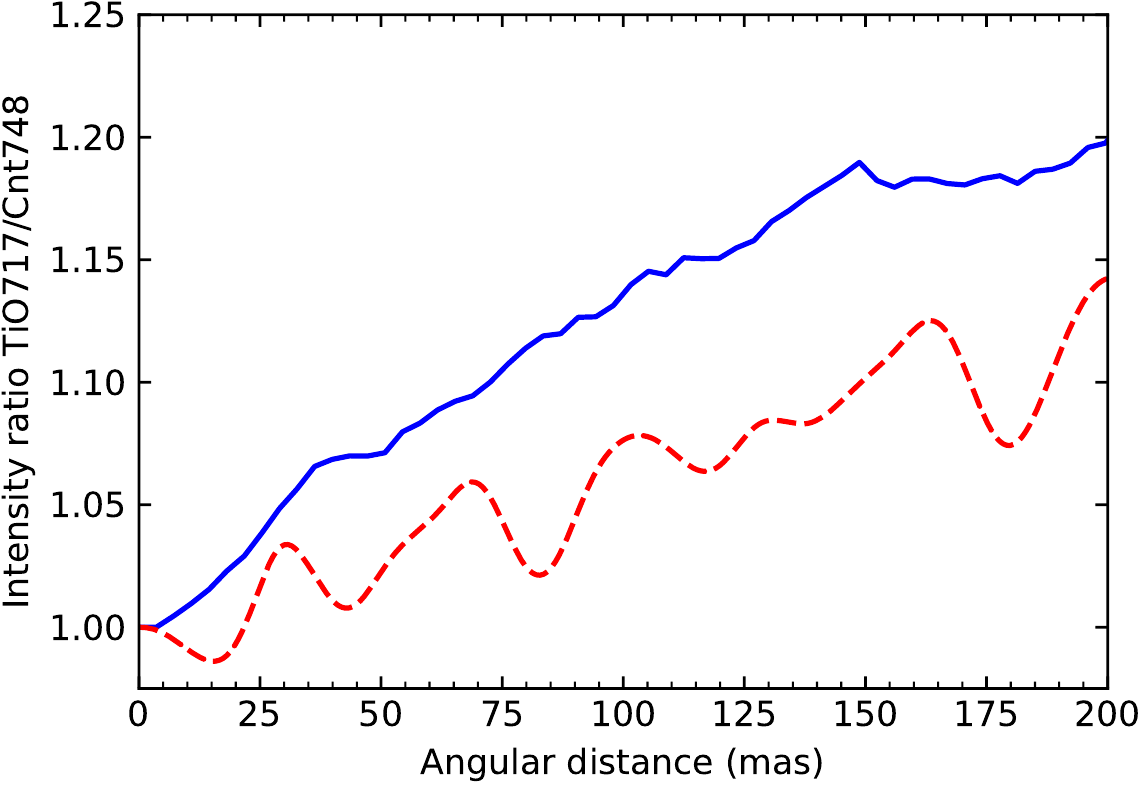}}}
                \end{center}
                \caption{Average 1-D line profile of  observations (solid blue line) and the best-fit model (dashed red line) measured from the ratio of the intensity maps taken at 717\,nm (TiO717) and 748\,nm (Cnt748).}
                \label{fig:ratios}
        \end{figure}
        
        To study the extended TiO emission of \ik, we subtracted the TiO717 and Cnt748 images normalized with the intensity at the centre of the stellar disk, following the approach adopted in \cite{2017A&A...597A..20O}. This method allows us to recognize the extended emission in positive pixel values and ascertain its spatial extent, but on the cost of the absolute flux scale, meaning with this method we cannot ascertain the exact morphology of the TiO emission observed over the central region.
        The right panel of Fig.~\ref{fig:cont_subtr} shows the continuum-subtracted TiO717 image. As can be seen, the TiO emission is concentrated in an apparently elongated area with an extension of about 80\,mas\,$\times$\,50\,mas (8\,$\times$\,5\,\rstar), which is also visible in the polarized intensity map at 717\,nm in Fig.~\ref{fig:1}. We note, however, that since the absolute flux scale is lost, we cannot address the morphology of the central region. 
        
        To examine whether or not the larger extension of the image at TiO717 can be explained solely by stronger dust-scattered light at Cnt717 than at Cnt748, we estimated the possible dust contribution using our best-fit model obtained from our 2-D radiative transfer modelling, presented in Sect.~\ref{sec:rtm}. In particular, we used the ratio of the TiO717 and Cnt748 filter. Figure~\ref{fig:ratios} shows the average 1-D line profile of these ratios measured for the observation (solid line) and the best-fit dust model (dashed line), respectively. Since the models in each filter are convolved with the corresponding PSF, the model images not only include the effects of dust, but also the imprint of the AO performance in each filter. As can be seen in Fig.~\ref{fig:ratios}, the apparent extended emission cannot be solely explained with dust, present in the CSE of \ik.

        The observed TiO emission extends in total to about 150\,mas ($\sim$15\,\rs). At this distance from the central star, however, the gas temperatures are fairly low ($\lesssim$\,$400$\,K, e.g. \cite{2010A&A...516A..69D}), which causes the thermal emission to be very weak at the observed wavelengths indicating that the detected emission originates from scattering on TiO gas in the extended atmosphere, instead of thermal emission from the gas.
        
        The observations are consistent with speckle interferometric observations in the TiO bands of \object{R~Leo} and \object{R~Cas} obtained by \citet{2001A&A...376..518H} and \citet{1996A&A...316L..21W}, respectively, as well as in general agreement with results determined for \ik\ by \cite{2018A&A...615A..28D} in their ALMA spectral-line and imaging survey. Their data with a spatial resolution of 150\,mas revealed that TiO and TiO$_2$ gas, among many others, are traceable up to $\sim$15\,\rs\ and even up to about 28\,\rs, respectively. This indicates that some fraction of the detected TiO and TiO$_2$ gas does not partake in dust nucleation and growth. The remaining gas is transported outwards into the CSE, which could partially explain the extremely broad intensity profile we observed (see Sect.~\ref{sec:rtm}). Their channel maps and total intensity maps show the presence of blobs in the inner wind region within 400\,mas. This non-homogeneous density distribution can result in a more active photochemistry in the inner winds, since lower density regions will allow energetic interstellar UV photons to penetrate deeper into the envelope. 
        
        \begin{figure*}
                \begin{center}
                        \resizebox{\hsize}{!}{\rotatebox{0}{\includegraphics{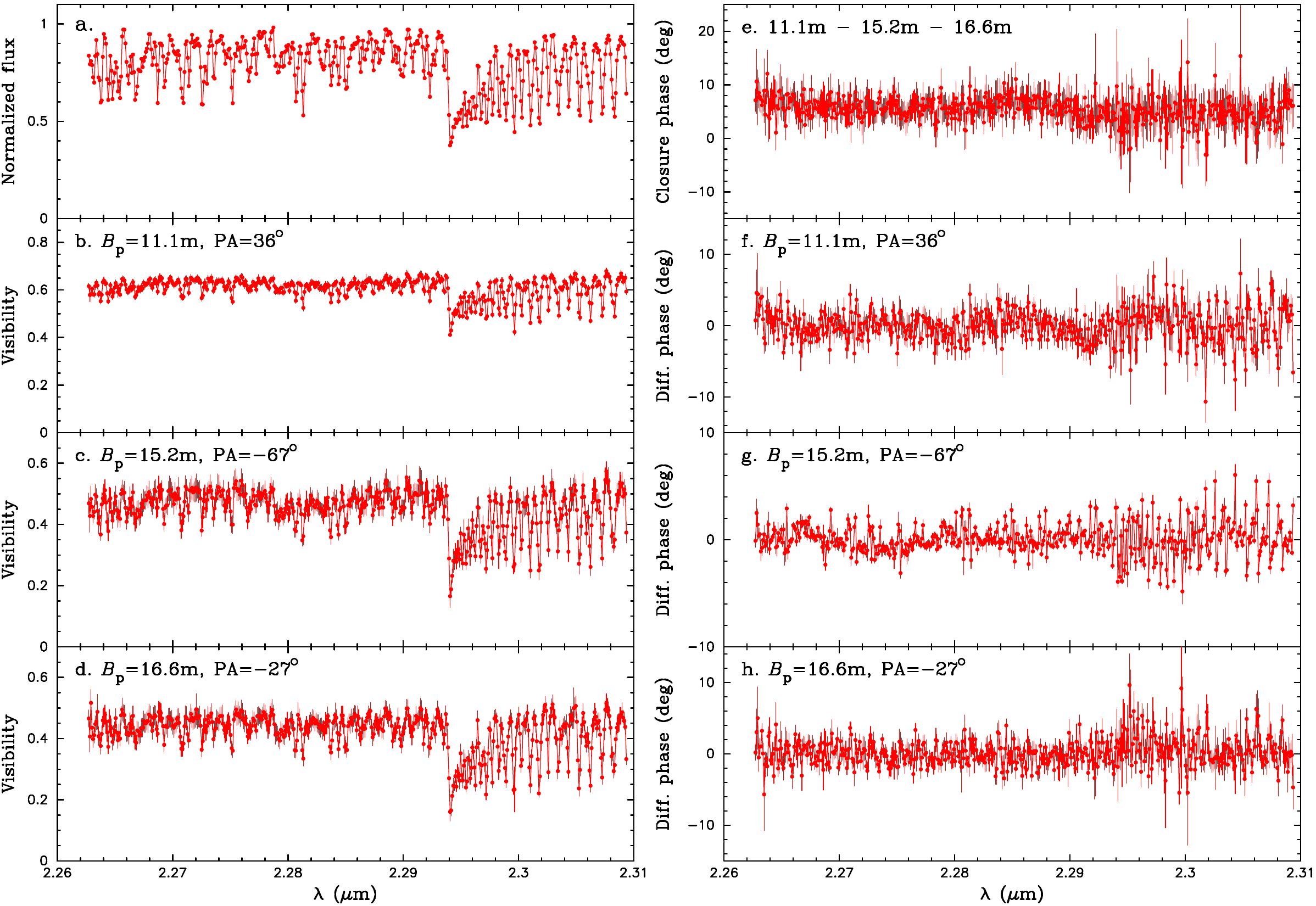}}}
                \end{center}
                \caption{
                        VLTI/AMBER observation of \ik\ with spectral resolution of 12\,000. 
                        {\bf a:} Observed spectrum. 
                        {\bf b}--{\bf d:} Visibilities. 
                        {\bf e:} Closure phase. 
                        {\bf f}--{\bf h:} Differential phases. 
                }
                \label{iktau_amber_res}
        \end{figure*}
        
        \subsection{AMBER spectro-interferometric observations of the central star}
        \label{subsect_res_amber}
        
        Figure~\ref{iktau_amber_res} shows the spectrum, visibilities, DPs, and CP measured with VLTI/AMBER. The visibilities show significant decreases in the first overtone lines of CO, which means that the star appears larger in the CO lines than in the 
        continuum. While the CP and the DP measured at the 16.6\,m baseline do not 
        clearly show non-zero values within the errors, the deviation from zero is 
        clearer in the DPs measured in the CO lines at the 11.1 and 15.2\,baselines. 
        This means asymmetry in the CO-line-forming outer atmosphere, as has often 
        been detected in the individual CO lines in other AGB stars \citep[e.g.][]{2012A&A...537A..53O,2016A&A...589A..91O}. 
        
\begin{figure}[!hbtp]
                \begin{center}
                        \resizebox{\hsize}{!}{\rotatebox{0}{\includegraphics{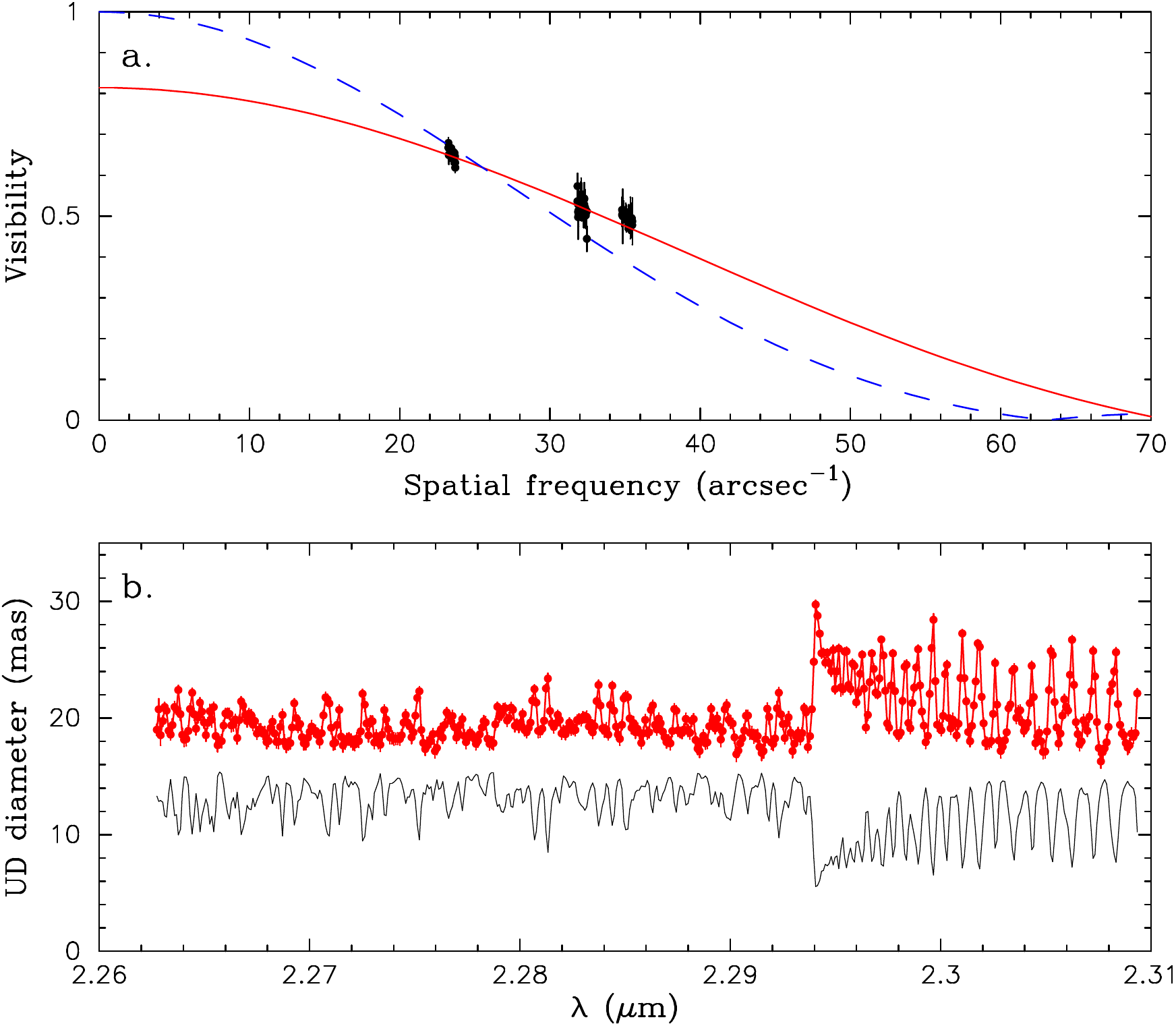}}}
                \end{center}
                \caption{
                        {\bf a:} Visibilities observed at continuum wavelengths are plotted with the black dots. The dashed blue line represents the power-law-type, limb-darkened disc fit. The solid red line shows the best-fit with a power-law-type, limb-darkened disc with an extended, resolved-out component. {\bf b:} Uniform-disc diameter across the observed wavelength range (solid red line). The observed, scaled spectrum is shown with the thin black 
                        line. 
                }
                \label{iktau_amber_lddudfit}
        \end{figure}

        To estimate the angular size of the star, we first fitted the observed visibilities 
        with a power-law-type, limb-darkened disc (LDD) \citep{1997A&A...327..199H}. We only used the visibilities measured in the continuum wavelength points, 
        avoiding the CO lines as well as the weak lines present shortwards of the 
        CO band head at 2.294\,\microns. The derived LDD diameter and the limb-darkening parameter are 37.00$\pm$0.79\,mas and 5.5$\pm$0.3, respectively. 
        However, as Fig.~\ref{iktau_amber_lddudfit}a (dashed blue line) shows, 
        the fit to the observed visibilities is poor (reduced $\chi^2$=6.1). 
        Moreover, the derived LDD diameter of 37.00\,mas results in an effective temperature as low as 1670\,K when combined with the observed bolometric flux presented in 
        Sect.~\ref{sec:stell_par}. 
        
        \cite{2004ApJ...605..436M} note that their aperture-masking data 
        of \ik\ obtained at 2.2\,\microns\ can be better explained by the central star and a very extended component that accounts for 28\% of the total flux at the observed wavelength. The extended component is likely a dust shell as modelled by the authors. 
        Therefore, we fitted our AMBER data with an LDD and an extended 
        component that is resolved out with our AMBER baselines. 
        As Fig.~\ref{iktau_amber_lddudfit}a (solid red line) shows, the fit is better 
        (reduced $\chi^2$=0.79), and the derived angular diameter of 
        20.70$\pm$1.53\,mas is consistent with the 20.2\,mas derived by \cite{2004ApJ...605..436M}. The limb-darkening parameter is 
        1.2$\pm$0.8, which is not well constrained because our AMBER baselines 
        are all on the first visibility lobe. The fractional flux contribution of the extended, resolved-out component is 0.19$\pm$0.0062. The difference in the 
        fractional flux of the extended component between \cite{2004ApJ...605..436M} and our work may be due to the time variation in the circumstellar envelope and/or the difference in the field of view of the observations. 
        
        To estimate the angular size of the star in the spectral lines, we assumed 
        that the fractional flux contribution of the extended, resolved-out 
        component does not change in the observed wavelength range. This is reasonable because the dust shell is likely responsible for the extended emission, and 
        dust emission does not change significantly across the observed narrow spectral 
        range from 2.26 to 2.31\,\microns. Then we fitted the visibilities observed at each wavelength with a uniform disc with the resolved-out component and the fractional flux contribution of 19\%. Figure~\ref{iktau_amber_lddudfit}b shows that the angular diameter of \ik\ increases up to 25--30~mas in the CO lines (30--50\% larger than in the continuum). The extended CO atmosphere has been detected 
        in \wh\ as well \citep[Fig.~5 of][]{2016A&A...589A..91O}, although it 
        is much more pronounced in \wh\ than in \ik. 
        
        The mass-loss rate of \ik\ is about 20--35 times higher than that of \wh\ \cite[\mloss\,$\sim$1.3--2.3$\times$10$^{-7}$\,\my,][]{2014A&A...561A...5K,2012A&A...546A..16R}. Also, the variability amplitude and pulsation period of \ik\ (up to 6\,mag over 460\,days) are higher and longer than those of \wh\ (3\,mag over 389\,days). The effective temperatures of both stars are approximately the same (see Sect.~\ref{sec:stell_par}). The luminosity of \ik\ is approximately twice as high as that of \wh\, while the mass is estimated to be 1--1.5\,\msun\ for both stars \citep{2017A&A...602A..14D}. Therefore, the surface gravity of \ik\ is nearly half that of \wh. All of it suggests that \ik\ should show a CO atmosphere more pronounced than \wh. Nevertheless, the AMBER observations show that it is not the case. It may be related to non-equilibrium chemistry affected by shocks, which may be stronger in \ik\ given its larger pulsation amplitude. However, the reason for the less pronounced CO atmosphere of \ik\ remains unclear at the moment.
        
        \begin{table}[hbtp!] 
                \caption{Stellar parameter of \ik.} 
                \label{table:4} 
                \centering      
                \begin{tabular}{l c c}
                        \hline\hline\\[-1em]
                        Parameter & Value & Reference \\
                        \hline 
                        $D$\,[pc] & 280$\pm$30 & a, b, c \\
                        \av\,[mag] & 0.351$\pm$0.172 & - \\
                        \rs\,[cm] & (4.23$\pm$0.46)\,$\times$\,10$^{13}$ & - \\
                        \fbol\,[$\mathrm{W\,m^{-2}}$] & (3.56$\pm$0.18)\,$\times$\,10$^{-9}$ & - \\
                        \lbol\,[\ls] & 8724$\pm$1921 & - \\
                        \teff\,[K]  & 2234$\pm$86 & - \\
                        $M^{\star}_{\mathrm{ZAMS}}$\,[$M_{\sun}$] & 1--1.5 & d\\
                        $\log g$ & -0.95 & - \\
                        \hline 
                \end{tabular}
                \tablefoot{$D$: Distance to the Sun. \av: Interstellar extinction in the V-band. $\theta_{\mathrm{D}}$: Angular diameter. \rs: Stellar radius. \fbol: Bolometric flux. \lbol: Bolometric luminosity. \teff: Effective temperature. $M^{\star}_{\mathrm{ZAMS}}$: Stellar mass at the zero age main sequence (ZAMS). $\log g$: Logarithm of the stellar surface acceleration. (a): \citet{1997ApJ...490..407H}; (b): \citet{2012A&A...546A..16R}; (c): \citet[Gaia DR2,][]{2018A&A...616A...1G}; (d): \citet{2017A&A...602A..14D} ; (-): this work.}
        \end{table}

        \section{Determination of stellar parameter} \label{sec:stell_par}
        
        \begin{figure}
                \begin{center}
                        \resizebox{\hsize}{!}{\rotatebox{0}{\includegraphics{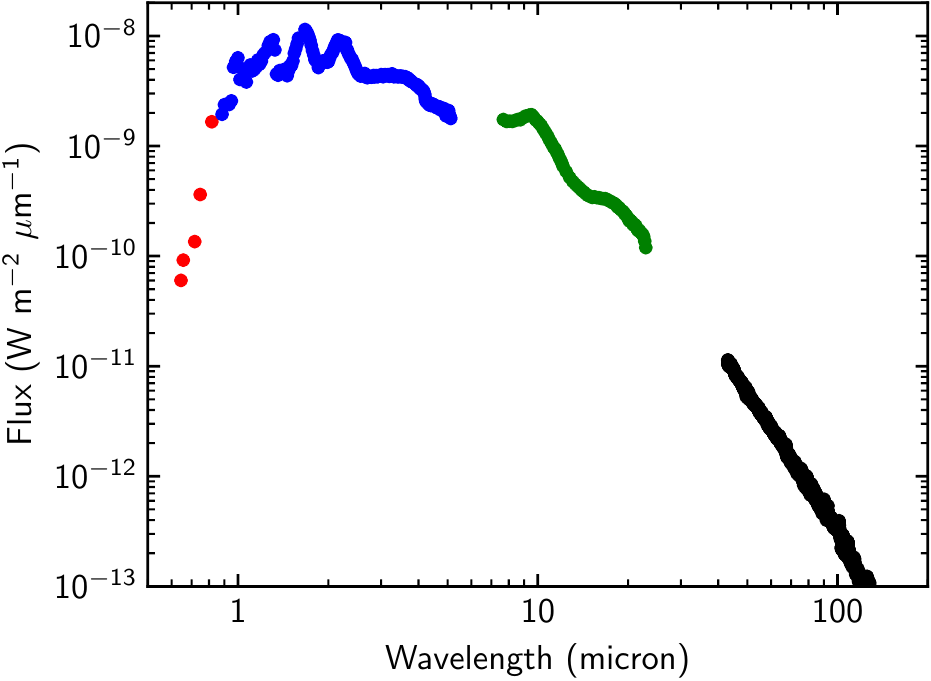}}}
                \end{center}
                \caption{Photometric and spectrophotometric data of \ik\ obtained from our SPHERE-ZIMPOL observations (red), the Cassini Atlas of Stellar Spectra (CAOSS, blue), the Infrared Astronomical Satellite (IRAS-LRS, green) and Long Wavelength Spectrometer (LWS) spectrum from the Infrared Space Observatory (ISO-LWS, black).}
                \label{fig:flux_iktau}
        \end{figure}

        The effective temperature (\teff) and bolometric luminosity (\lbol) corresponding to the variability phase of our SPHERE-ZIMPOL observations are needed for our radiative transfer modelling. We use our flux-calibrated SPHERE-ZIMPOL observations as well as spectro-photometric data available in the literature to estimate the bolometric flux (\fbol) at the time of observation (phase\,$=$\,0.27 post-maximum light).
        In the visible wavelength range (645--820\,nm) we used the fluxes derived with the five SPHERE-ZIMPOL filters (Table~\ref{table:2}).
        In the near-infrared, between 0.8\,\microns\ and 5.1\,\microns, we used data from the Cassini Atlas of Stellar Spectra \citep[CAOSS,][]{2015ApJS..221...30S} observed on 2002 July 19. This date corresponds to phase 0.26 at pre-maximum light applying the phase relation derived by \cite{2018A&A...612A..48W}. 
        For the mid- and far-infrared regions we obtained low resolution spectra (LRS) from the Infrared Astronomical Satellite \citep[IRAS-LRS,][]{1986A&AS...65..607O} from 8\,\microns\ to 25\,\microns, and the Infrared Space Observatory \citep[ISO,][]{1996A&A...315L..27K} Long-Wavelength Spectrometer (LWS) covering 43\,\microns\ to 196\,\microns. The resulting spectral energy distribution is shown in Fig.~\ref{fig:flux_iktau}.
        We then de-reddened all spectro-photometric data using the wavelength-dependence derived by \cite{1989ApJ...345..245C} and \av=0.351$\pm$0.172\,mag, which we derived from the parametrization of the interstellar extinction \citep{1992A&A...258..104A}.
        The integration of the resulting de-reddened fluxes over the available wavelength range yields \fbol=(3.56$\pm$0.18)\,$\times$\,10$^{-9}\, \mathrm{W\,m^{-2}}$. This error estimate only accounts for the variation of the interstellar extinction, due to the uncertainty in the distance of \ik.
        The result, however, is in broad agreement with other flux estimates such as \fbol=(3.31$\pm$1.10)\,$\times$\,10$^{-9}\, \mathrm{W\,m^{-2}}$ by \cite{2016AJ....152...16V}, who performed spectral energy distribution (SED) fitting on photometric data of \ik\ obtained by \cite{1974ApJ...189...89D} on 1971 November 21, which corresponds to phase $\sim$0.9 (pre-maximum light). From the distance of $D$=280$\pm$30\,pc, the angular diameter $\theta_{\mathrm{D}}$=20.7$\pm$1.5\,mas derived from our VLTI/AMBER observations and our flux estimate, we compute a bolometric luminosity of \lbol=8724$\pm$1921\,\ls\ and an effective temperature \teff=2234$\pm$86\,K for \ik\ at the date of our observation, which we applied to our model. The derived stellar parameters are summarized in Table~\ref{table:4}.

        \begin{figure}[hbtp!] 
                \begin{center}
                        \resizebox{40mm}{!}{\rotatebox{0}{\includegraphics{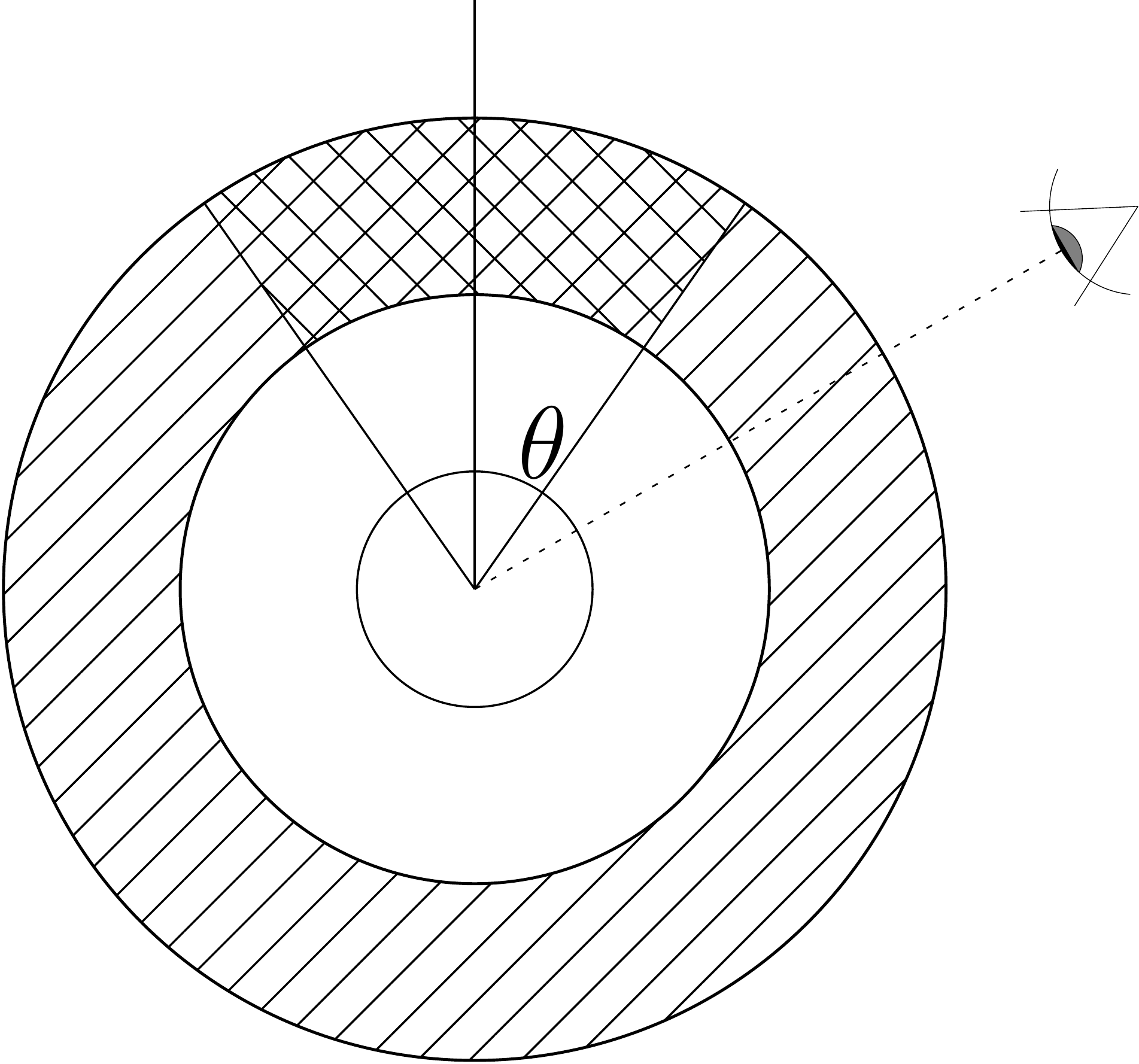}}}
                \end{center}
                \caption{Schematic view of our 2-D dust clump model \cite[reproduced from][]{2016A&A...589A..91O}. The hatched region represents the spherical dust shell, while the cross-hatched region shows the cone-shaped density enhancement, which is characterized by the half-opening angle $\theta$. A viewing angle of 60\degr\ is shown as an example.}
                \label{fig:rtm_scheme}
        \end{figure}
        
        \section{Monte Carlo radiative transfer modelling}      \label{sec:rtm}
                
        For the interpretation of our SPHERE-ZIMPOL observations of \ik\ and in order to constrain the physical properties of the dust-forming regions close to the star, we used the 2-D Monte Carlo radiative transfer code mcsim\_mpi \citep{2006A&A...445.1015O}, which has already been applied to polarimetric images of the AGB star \wh\ \citep{2016A&A...589A..91O,2017A&A...597A..20O}.
        

        As described in Sect.~\ref{sec:obs_res}, we detected two distinct dust clumps extending towards the NE and the south, both very similar in regards to their polarization properties, for example in their extent or their degree of linear polarization. We therefore primarily focused on the reproduction of one clump, namely the main clump NE, instead of reproducing the entire image. As a result, the model set-up is almost the same used to model the dust clumps detected in \wh\ by \cite{2016A&A...589A..91O,2017A&A...597A..20O}. 
        
        To model our observations, we adopt a simple single-shell model with a cone-shaped density enhancement which is described by the density ratio within and outside the cone. The cone, shown in Fig.~\ref{fig:rtm_scheme}, is characterized by the half-opening angle $\theta$\,=\,40\degr. 
        We estimated this angle from the north-eastern dust clump in the polarized intensity images as well as from the RMS maps where we measured the angle from the boundary of the area where $\Delta I_{\mathrm{p}}/I_{\mathrm{p}}$$\leq$10\,\%. 
        
        The free parameters of our model are the optical depth in the visible (550\,nm) of the dust shell in radial direction, its inner and outer boundary radius, the exponent of the radial density distribution ($\propto r^{-p}$), and the density ratio between the cone and the remaining shell. The dust opacities are treated with the code developed by \cite{1983asls.book.....B}, which we used to compute the scattering and absorption coefficients, as well as the scattering matrix elements from complex refractive indices of \alum\ measured by \citet[][their `Alumina' sample]{1995Icar..114..203K}, \enst, and \fors\ measured by \citet{2003A&A...408..193J}. To compare the output of the simulation with our observations, the model $Q$ and $U$ images, as well as the total intensity $I$ were convolved with the observed PSF of the PSF reference star \hip\ before computing the maps of the polarized intensity and the degree of linear polarization as described in Sect.~\ref{sec:obs}. We used this approach rather than convolving the model $I_{\mathrm{p}}$ and $p_{\mathrm{L}}$ because it corresponds to how the $I_{\mathrm{p}}$ and $p_{\mathrm{L}}$ maps were obtained from the observational data. We did not consider the images taken at 656.3\,nm and 717\,nm due to the possible effects of the \hal\ and TiO emission (see Sect.~\ref{subsec:emmission}).
        
        \begin{figure*}[hbtp!] 
                
                \includegraphics[width=180mm]{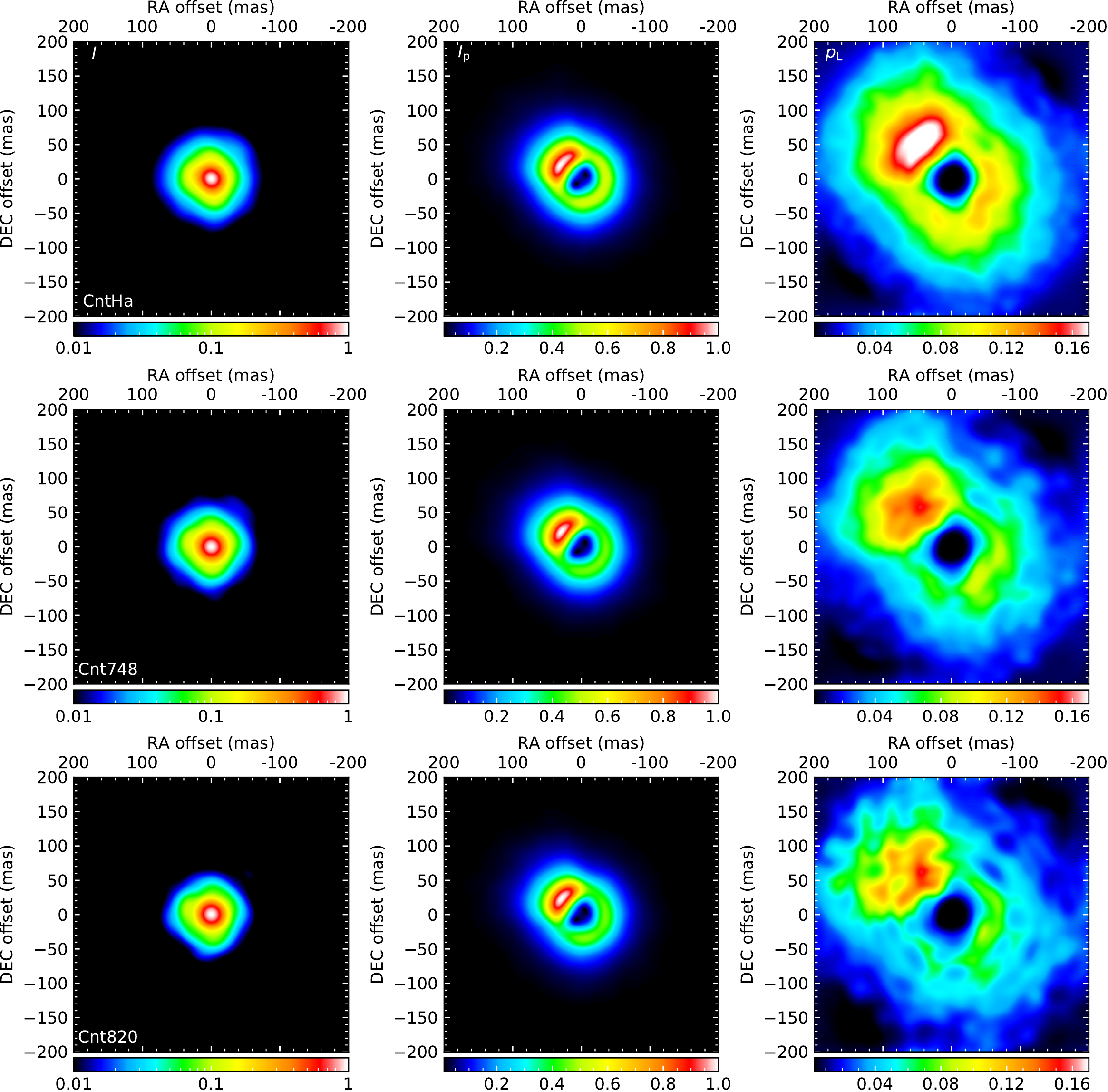}
                
                \caption{Best-fit dust clump model for \ik. Shown from left to right are the maps of the intensity $I$, polarized intensity $I_{\mathrm{p}}$, and degree of linear polarization $p_{\mathrm{L}}$ predicted by the model after convolution with the observed PSF. The model images predicted at 645\,nm, 748\,nm, and 820\,nm are presented in the columns from top to bottom, respectively. North is to the top, east to the left.
                }
                \label{fig:rtm_imgs}
        \end{figure*}
        
        \begin{figure*}[hbtp!] 
                \includegraphics[width=180mm]{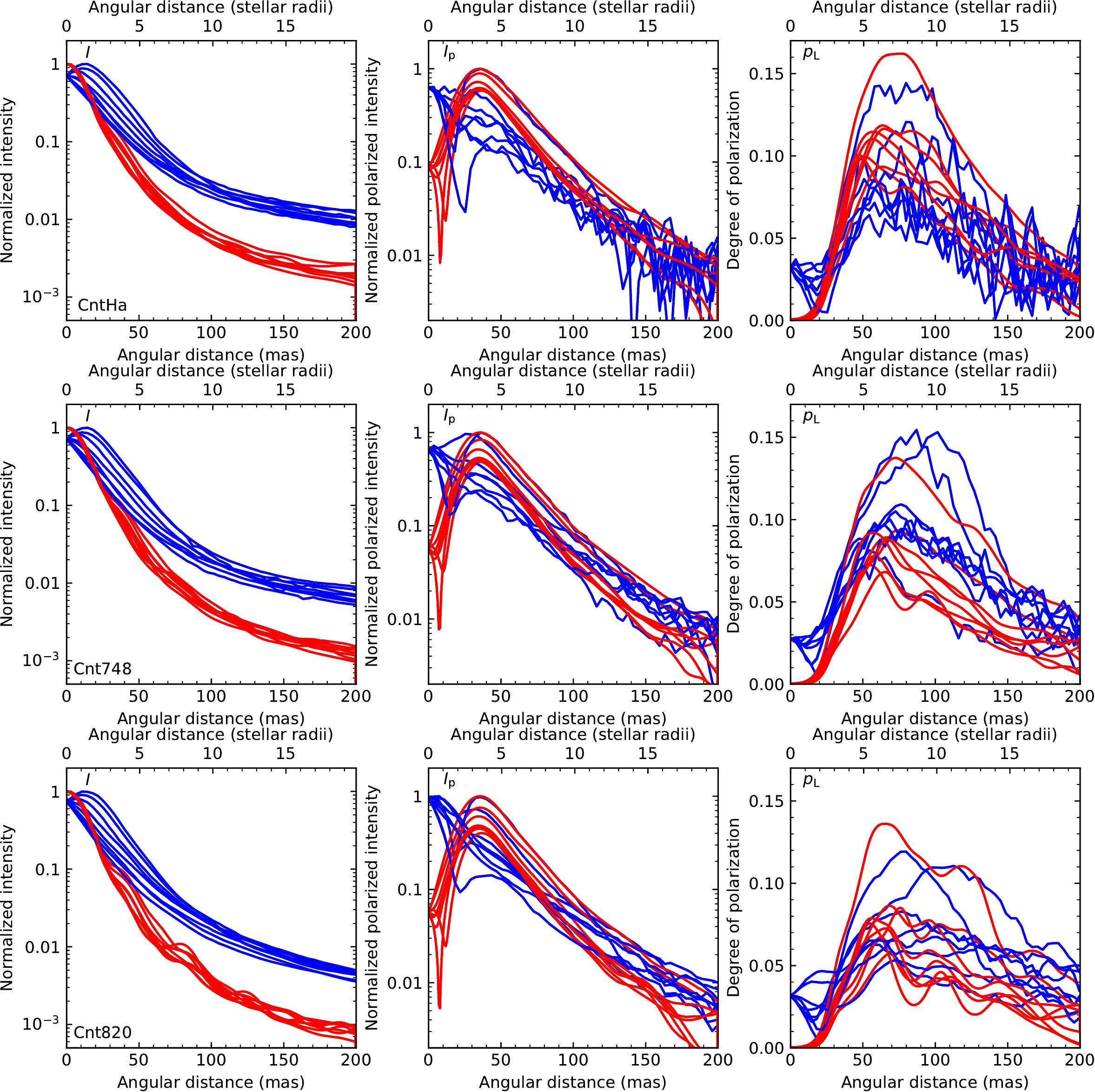}
                \caption{Comparison of  dust clump model and  observed SPHERE-ZIMPOL data. The first, second, and third columns show the 1-D cuts at eight different position angles (0\degr, 45\degr, 90\degr,..., 315\degr) of the intensity $I$, the normalized polarized intensity $I_{\mathrm{p}}$, and the degree of polarization $p_{\mathrm{L}}$ maps, respectively. The top, middle, and bottom rows show the comparison at 645\,nm, 748\,nm, and 820\,nm, respectively. In each panel, the observed data are plotted with blue lines, while the model is plotted with red lines.   
                }
                \label{fig:rtm_profs}
        \end{figure*}
        
        \begin{figure*}[hbtp!] 
                \includegraphics[width=180mm]{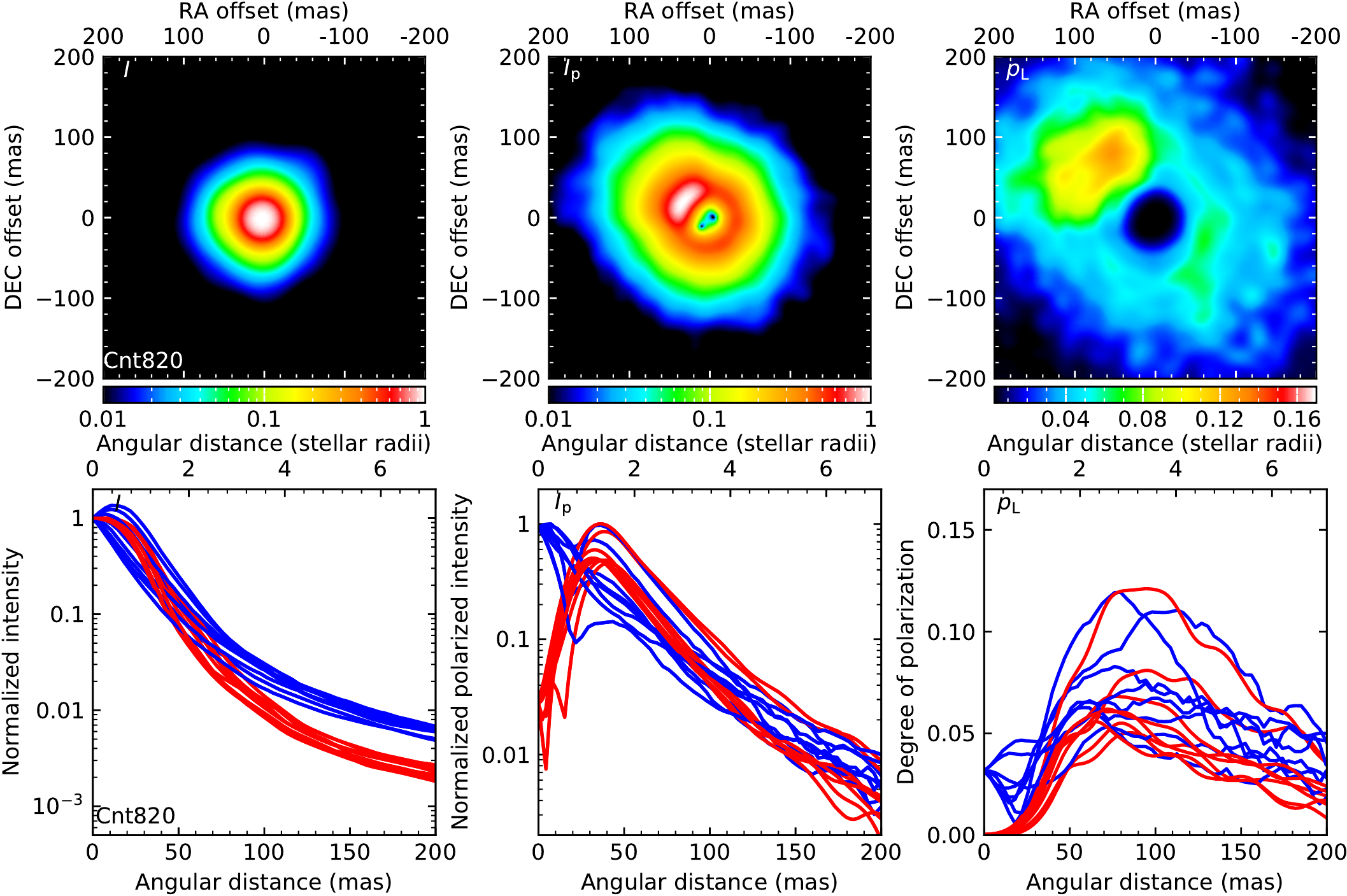}
                \caption{Best-fit dust clump model for \ik\ simulated at 820\,nm for a larger model star with radius \rsp. The top row shows from left to right the maps of the intensity $I$, polarized intensity $I_{\mathrm{p}}$, and degree of linear polarization $p_{\mathrm{L}}$ predicted by the model after convolution with the observed PSF. North is to the top, east to the left. The second row displays the comparison of the dust clump model and the observed SPHERE-ZIMPOL data showing the 1-D cuts at eight different position angles (0\degr, 45\degr, 90\degr,..., 315\degr) of the intensity $I$, the normalized polarized intensity $I_{\mathrm{p}}$, and degree of polarization $p_{\mathrm{L}}$ maps, respectively. In each panel, the observed data are plotted with blue lines, while the model is plotted with red lines.}
                \label{fig:rtm_profs2}
        \end{figure*}
        
        We first approximated the radiation of the central star with a black body with an effective temperature of \teff=2234\,K. Figure~\ref{fig:rtm_imgs} shows the total intensity, polarized intensity, and degree of linear polarization maps at 645\,nm, 748\,nm, and 820\,nm obtained from our best-fit model. The best-fit model is characterized by \alum\ grains with a size of 0.1\,\microns, a density enhancement in a single shell with half-opening angle 40\degr$\pm$10\degr, a density distribution power-law exponent of 2.7, and a density ratio of 3.0$\pm$0.5 within and outside the cone. The inner boundary radius is 3.5$\pm$0.5\,\rs, while the outer radius is $\geq$25\,\rs. The optical depth at 550\,nm is 0.5$\pm$0.1 and the viewing angle measured from the symmetry axis (see Fig.~\ref{fig:rtm_scheme}) is 60\degr. The dust mass of this model is 4.8\,$\times$\,10$^{-8}$\,\msun\ adopting a bulk density of 4\,\bdens\ for \alum. We note that the grainy structure visible in the degree of linear polarization is a result of the convolution of the model with the observed PSF.
        In Figure~\ref{fig:rtm_profs}, we show the 1-D profiles of the observed and model images measured at eight different position angles (0\degr, 45\degr, 90\degr,..., 315\degr). The centre of the images was assumed to be the centroid of the low-polarization region in the centre of the degree of polarization maps.
        
        We found that for all sampled wavelengths, the grain sizes both smaller and larger than 0.1\,\microns\ in general predict the degree of polarization to be too low compared to what we have observed. The size of grains smaller than 0.1\,\microns\ is due to the much smaller albedo of these grains, which significantly reduces the proportion of scattered light and increases the fraction of un-polarized direct starlight. We obtain very similar parameters for the models with \fors\ and \enst. Thus, we cannot distinguish between those three dust species from our SPHERE-ZIMPOL data alone. 

        
        The shape of polarized intensity and absolute values of the degree of polarization reasonably agree with the observed data, given the simplistic assumptions of the model and the fact that we considered a density enhancement only in one direction. As can be seen in the first column of Fig.~\ref{fig:rtm_profs}, however, the observed total intensity maps at the modelled wavelengths are significantly wider than those predicted by the model.
        
        We also considered models with multiple dust grain sizes. However, to keep the model simple we set the inner and outer boundary radii, as well as the density exponent and density enhancement to be the same for the two-grain species, meaning a combination of 0.1\,\microns\ and 0.3\,\microns\ \alum\ grains, while exploring different values for the optical depth at 550\,nm. We found that the normalized polarized intensity as well as the degree of polarization could also be explained to a fairly
reasonable level with models computed for \alum\ with grain sizes of 0.1\,\microns\ and 0.3\,\microns, and \tauV=0.7 (0.1\,\microns\ grains) and \tauV$\sim$\,0.1 (0.3\,\microns\ grains). These optical depths correspond to a dust mass of 4.7\,$\times$\,10$^{-8}$\,\msun\ for 0.1\,\microns\ grains and 3.7\,$\times$\,10$^{-9}$\,\msun\ for 0.3\,\microns\ grains. The two-grain models, however, cannot simultaneously explain the observed total intensity, polarized intensity, and degree of linear polarization, either. The changes in the fit to the observed data are marginal compared to our best single-grain model. Therefore, we can neither confirm nor reject the presence of grains larger than 0.1\,\microns, although, if present, these larger grains must account for a small fraction of the dust mass.
        
        As mentioned above, we approximated the effective temperature of the model star employing the stellar diameter measured with AMBER in the infrared. Given, however, that the spectra of oxygen-rich AGB stars in the visible wavelengths are often dominated by molecular bands, the actual spectrum of the real star might vary significantly from the black body spectrum, also causing the radius of the layer (whose radiation the dust grains can see) to vary significantly from wavelength to wavelength. To get an estimate of the effects of molecular bands we repeated our radiative transfer modelling, considering a larger and thus cooler model star. For simplicity we focus on our observations at 820\,nm, since we expect the brightness temperature (\tb) to vary between the different wavelengths of observation. 
        We first attempted to fit the observed total intensity and the observed flux at 820\,nm treating the model star radius (\rsp) and the temperature as free parameters. We note that in this case the model star radius (\rsp) refers to the radius of the layer from which the radiation seen by the dust grains seems to originate, while the temperature refers to the brightness temperature (\tb) of this layer, respectively. The observed total intensity is best reproduced assuming a model star radius of about 28\,mas whereas, on the other hand, the observed flux at 820\,nm (see Table~\ref{table:2}) is best reproduced by a brightness temperature (\tb) of about 1500\,K.
        We then re-adjusted the remaining model parameters to also fit the polarized intensity and the degree of linear polarization.
        
        Figure~\ref{fig:rtm_profs2} shows the best fit to the total intensity, polarized intensity, and degree of linear polarization maps at 820\,nm, as well as the 1-D profiles of the observed and model images measured at eight different position angles (0\degr, 45\degr, 90\degr,..., 315\degr) obtained from our new model. As mentioned, we find that the shape of the total and the polarized intensity, as well as the degree of polarization being reasonably reproduced by a model star with an inner boundary radius of about 28\,mas (1\,\rsp), an outer radius of 250\,mas ($\geq$9\,\rsp), as well as an optical depth at 550\,nm of about 0.8, and a density distribution exponent of 2.8, respectively. The dust mass of this model is about 5.2\,$\times$\,10$^{-8}$\,\msun\ adopting a bulk density of 4\,\bdens\ for \alum.
        
        A thorough modelling with the effects of the molecular bands fully taken into account remains difficult,however, as the comparison of both models also shows that the effects of molecular bands alone only allow to partially explain the observed total intensity. A more realistic atmospheric model is required in order to fully understand the effects of the molecular bands on the dust formation at different wavelengths.
        
        Another reason for the observed differences in the total intensity in the investigated parameter space, therefore, might be the difference in the Strehl ratio seen in the observations of \ik\ and the PSF reference star. Given the higher Strehl ratio observed for the PSF reference, its profile is narrower than that of \ik. This could partially explain the narrower intensity profile of the model, since we used the PSF image to convolve the model images. However, while the Strehl ratios for \ik\ are comparable to those for the PSF reference, it is unlikely that the discrepancy between the data and the models can be explained by the differences in the Strehl ratios alone. The discrepancy might also be attributed to the extended TiO emission described in Sect.~\ref{subsec:emmission}. Because the visible light is dominated by the TiO bands, which in turn can broaden the intensity profile as well. Another possibility is the applied grain model. Since our model only includes spherical dust grains we cannot estimate the influence on the results if instead hollow spheres were used, as employed in the modelling of \object{R~Dor} by \cite{2016A&A...591A..70K}.
        
        \citet[][and references therein]{2016A&A...585A...6G} have presented comprehensive models for non-equilibrium chemical processes of gas and dust in the inner wind of \ik, incorporating the nucleation of dust grains from the gas phase. Their models predict the formation of \alum\ close to the star ($\leq$2\,\rs), as well as the formation of silicate dust at radii larger than 3.5\,\rs\ This suggests that the passage of shocks destroys the grains completely, and small grains start to form only half a period after the passage of the shocks, followed by efficient grain growth. Our model suggests the inner radius of the dust shell of $\sim$3.5\,\rs. Therefore, the grain responsible for the scattered light may be iron-free silicates. Alternatively, as suggested by \cite{1997Ap&SS.255..437K,1997Ap&SS.251..165K} and \cite{2016A&A...594A.108H}, the responsible grain may be \alum\ with iron-free silicate mantles, although our data and modelling do not allow us to distinguish \alum\ and iron-free silicates \fors\ and \enst\ as already mentioned.

        \section{Conclusion} \label{sec:concl}
        
        We have presented visible polarimetric imaging observations of the oxygen-rich AGB star \ik\ obtained with SPHERE-ZIMPOL at post-maximum light (phase 0.27). The polarized intensity maps taken with SPHERE-ZIMPOL at five wavelengths from 645 to 820\,nm and a spatial resolutions of 20--30\,mas show two distinct clumpy dust clouds within 50\,mas (5\,\rs) of the central star. The observed images furthermore reveal diffuse clouds extending towards the north and west up to about 500\,mas ($\sim$50\,\rs), as well as a dust cloud extending towards the south-east reaching as far as 73\,\rs. 
        
%
        The polarimetric imaging observations of \ik\ have revealed for the first time dust clouds forming close to the star in a high mass-loss AGB star with mass-loss rates 20 to 50 times higher than that of objects, such as \wh, \object{R~Dor}, and \object{$o$~Cet}, previously studied in a similar way. Moreover, the dust formation appears to be clumpy, which is also very similar to observations of AGB stars with lower mass-loss rates.
        
        Our 2-D Monte Carlo radiative transfer modelling shows that the observed degree of polarization and the normalized polarized intensity can be explained by an optically thin (\tauV=0.5$\pm$0.1) shell with $\sim$0.1\,\microns\ grains of \alum, \fors, or \enst, and an inner and outer radius of 3.5$\pm$0.5\,\rs\ and $\gtrsim$25\,\rs. Our model, however, cannot explain the observed total intensity of \ik.
        
        Given the periodic variations in brightness seen in the light curve of \ik\, and the fact that our observations took place at post-maximum light, monitoring observations of the variations in the clumpy dust clouds and of the extended \hal\ and TiO emissions following the variability phase of \ik\, are needed to understand the role of pulsation-induced shocks in dust and molecule formation and, consequently mass loss. Polarimetric imaging in the near-infrared, where the effects of the TiO bands are negligible, will help to further constrain the dust properties of the innermost regions of \ik.

        \begin{acknowledgements}
                We thank the ESO Paranal team for supporting our SPHERE and AMBER observations. C. A. acknowledges the financial support from Comit\'{e} Mixto ESO--Gobierno de Chile. K. O. acknowledges the support of the Comisi\'{o}n Nacional de Investigaci\'{o}n Cient\'{i}fica y Tecnol\'{o}gica (CONICYT) through the FONDECYT Regular grant 1180066. This research has made use of the SIMBAD database, operated at CDS, Strasbourg, France. We acknowledge with thanks the variable star observations from the AAVSO International Database contributed by observers worldwide and used in this research. This work has made use of data from the European Space Agency (ESA) mission
                {\it Gaia} (\url{https://www.cosmos.esa.int/gaia}), processed by the {\it Gaia}
                Data Processing and Analysis Consortium (DPAC,
                \url{https://www.cosmos.esa.int/web/gaia/dpac/consortium}). Funding for the DPAC has been provided by national institutions, in particular the institutions participating in the {\it Gaia} Multilateral Agreement. 
                
        \end{acknowledgements}
        
        %
        %
        
        \bibliographystyle{aa}
        \bibliography{iktau_adam_references.bib}
        
        \appendix

\end{document}